\documentclass[twocolumn]{aastex61}
\usepackage{amsmath,amssymb,amsthm}
\usepackage{wasysym}
\usepackage{bm}
\usepackage{parskip}
\usepackage{tikz}
\usetikzlibrary{arrows.meta}
\usepackage[normalem]{ulem}

\usepackage{graphicx}                 
\usepackage{xcolor}
\usepackage{tikz}
\usetikzlibrary{fpu,calc,intersections,through,backgrounds}
\usepackage{wrapfig}
\usepackage{MnSymbol}

\received{February 19, 2021}
\revised{April 20, 2021}
\accepted{\today}
\submitjournal{AJ}

\begin{document}

\title{Limiting behavior of asteroid obliquity and spin using a semi-analytic thermal model of the YORP effect}

\correspondingauthor{Oleksiy Golubov}
\email{oleksiy.golubov@karazin.ua}

\author[0000-0002-2427-9101]{Oleksiy Golubov}
\affiliation{Institute of Astronomy of V. N. Karazin Kharkiv National University, 35 Sumska Str., Kharkiv, 61022, Ukraine}

\author[0000-0003-0459-999X]{Vladyslav Unukovych}
\affiliation{Institute of Physics and Technology of V. N. Karazin Kharkiv National University, 4 Svobody Sq., Kharkiv, 61022, Ukraine}

\author[0000-0003-0558-3842]{Daniel J. Scheeres}
\affiliation{Department of Aerospace Engineering Sciences, University of Colorado at Boulder \\
	429 UCB, Boulder, CO, 80309, USA}

\begin{abstract}
	The Yarkovsky--O'Keefe--Radzievskii--Paddack (YORP) effect governs the spin evolution of small asteroids. The axial component of YORP, which alters the rotation rate of the asteroid, is mostly independent of its thermal inertia, while the obliquity component is very sensitive to  the thermal model of the asteroid.
	
	Here we develop a semi-analytic theory for the obliquity component of YORP. We integrate an approximate thermal model over the surface of an asteroid, and find an analytic expression for the obliquity component in terms of two YORP coefficients.
	
	This approach allows us to investigate the overall evolution of asteroid rotation state, and to generalize the results previously obtained in the case of zero thermal inertia. 
	
	The proposed theory also explains how a non-zero obliquity component of YORP originates even for a symmetric asteroid due to its finite thermal inertia. In many cases, this causes equatorial planes of asteroids to align with their orbital planes.
	
	The studied non-trivial behavior of YORP as a function of thermal model allows for a new kind of rotational equilibria, which can have important evolutionary consequences for asteroids.
\end{abstract}


\section{Introduction}

Rotation of kilometer-sized asteroids is governed by the YORP effect 
\citep{rubincam00,vokrouhlicky15}.
It is a torque caused by scattering and re-emission of light by asteroid's surface,
which can change both the asteroid's rotation rate $\omega$ and the obliquity $\varepsilon$. The part of the torque changing the rotation rate is called the axial component $T_z$, whereas the part affecting obliquity is called the obliquity component $T_\varepsilon$. Characterizing the overall evolution of the rotation state of an asteroid under the combined action of these two components is one of the most fundamental tasks of the YORP theory.

This task has already been solved in our previous paper in a simplified case of zero thermal inertia \cite{golubov19}. Under this constraining assumption, the evolution of asteroids has been simulated numerically, as well as studied analytically in the most typical case. It was shown that most of the asteroids when starting their evolution from slow rotation rate, gradually increase it to a certain limit, and if not getting disrupted by the centrifugal forces in the process, return back to very slow rotation. Inclusion of the tangential YORP into the model \cite{golubov12} can qualitatively alter this typical evolution and bring in the possibility of YORP equilibria.

Still, as long as the tangential YORP is disregarded and only the normal YORP is considered, the axial component is indeed to a very high accuracy independent of the thermal model. This fundamental fact about YORP was demonstrated in the simulations by \cite{bbc10} and later theoretically proven under more general assumptions by \cite{golubov16nyorp}. It allows to study $T_z$ in the limit of zero thermal inertia, as it is done in the model by \cite{golubov19}.

On the other hand, the obliquity component $T_\varepsilon$ is highly sensitive to the thermal inertia. As it has been shown in the numeric simulations by \cite{capek04}, $T_\varepsilon$ can be dramatically altered and even flip sign with the change of the heat conductivity. The authors conducted a deep analysis of the YORP torques, but did not extend their formalism to study the overall asteroid evolution.

A more advanced evolutionary study was later performed by \cite{scheeres08}, although in a simplified model. Their approach avoided rigorous solution of the heat equation by introducing a fixed time lag between the absorption and emission of energy. This allowed the authors to characterize the rotational dynamics and in particular to find the possibility of stable equilibria.

Here we generalize the results of \cite{golubov19} for the case of non-zero thermal inertia, basing our approach on the formalism of \cite{golubov16nyorp}. It makes our theory more precise than \cite{scheeres08}, and allows to go father unto analysis of the asteroid than \cite{capek04}.

In Section \ref{sec:YORP coefficients} we combine the analytic and numeric approaches to simplify the problem and to reduce all the information about the asteroid shape to two YORP coefficients. The following Section \ref{sec:YORP eqilibria} studies the asteroid evolution in terms of the YORP coefficients, as well as the stable YORP equilibria that can arise on the evolutionary tracks of some asteroids. 

\section{YORP coefficients}
\label{sec:YORP coefficients}
\subsection{Problem setting}

Let us start with the equations of motion of the asteroid, which describes the evolution of its rotation rate $\omega$ and obliquity $\varepsilon$ as a function of time $t'$ \cite{rubincam00}:
\begin{eqnarray}
I_z \frac{\mathrm{d}\omega}{\mathrm{d}t'}&=&T_z ,\\
I_z \frac{\mathrm{d}\varepsilon}{\mathrm{d}t'}&=&\frac{1}{\omega}T_\varepsilon,
\label{dimensional_evolution_equations}
\end{eqnarray}
where $I_z$ is the asteroid's moment of inertia, while $T_z$ and $T_\varepsilon$ are the axial and obliquity components of the YORP torque, acting on the asteroid.

It is convenient to non-dimensionalize the problem in the following manner.
Let the mean volumetric radius of the asteroid be $R$ and its density $\rho$.
Then we can introduce the dimensionless moment of inertia $i_z$ by the following equation:
\begin{equation}
i_z = \frac{I_z}{\rho R^5}.
\end{equation}
The dimensionless YORP torques are introduced as
\begin{eqnarray}
\tau_z &=& \frac{cT_z}{\Phi R^3},\\
\tau_\varepsilon &=& \frac{cT_\varepsilon}{\Phi R^3}.
\label{nondimentionalization_T}
\end{eqnarray}
Here $c$ the speed of light and $\Phi$ is the effective solar constant.

Next it is convenient to introduce the dimensionless thermal parameter $\theta$,
\begin{equation}
\theta = \frac{\left(C\rho\omega\kappa\right)^{1/2}}{\left(\epsilon\sigma\right)^{1/4}\left(1-A\right)^{3/4}\Phi^{3/4}}.
\label{theta_definition}
\end{equation}
Here $A$ as the albedo, $\epsilon$ is the thermal emissivity, $\sigma$ is the Stefan--Boltzmann constant,
$\kappa$ is the heat conductivity of the material constituting the asteroid surface,
and $C$ is its specific heat capacity.
This thermal parameter characterizes the relative importance of thermal inertia of the surface:
for $\theta \ll 1$ the surface almost instantly adjusts its temperature to the illumination,
while for $\theta \gg 1$ the surface temperature remains almost constant throughout the rotation period.

The dimensionless time is introduced as $t=t'/t_0$, where
\begin{equation}
t_0 = \frac{\sqrt{\epsilon\sigma(1-A)^3\Phi}R^2c}{C\kappa}.
\label{t0_definition}
\end{equation}
The value of $t_0$ characterizes the order of magnitude of the YORP evolution timescale for the thermal parameter $\theta\sim 1$ and for the maximal possible strength of YORP $\tau\sim 1$. For other values of $\theta$ and $\tau$, the timescale would change in the direct proportion to $\theta^2\tau^{-1}$.

After all these changes of notation applied, Eqs. (\ref{dimensional_evolution_equations}) assume the following form,
\begin{eqnarray}
i_z \frac{\mathrm{d}\theta^2}{\mathrm{d}t}&=&\tau_z,
\label{dimensionless_evolution_equation1}\\
i_z \theta^2\frac{\mathrm{d}\varepsilon}{\mathrm{d}t}&=&\tau_\varepsilon.
\label{dimensionless_evolution_equation2}
\end{eqnarray}

\subsection{Obliquity component in terms of the YORP coefficients}
\label{sec:Obliquity component in terms of the YORP coefficients}

The dimensionless YORP torques $\tau_z$ and $\tau_\varepsilon$ can be expressed as integrals over the surface of the asteroid, containing dimensionless pressures $p^\tau_\mathrm{sin}$ and $p^\tau_\mathrm{cos}$. These pressures are some known functions of the thermal parameter $\theta$, obliquity $\varepsilon$, and the latitude of the point on the asteroid surface $\psi$. (See \cite{golubov16nyorp} for derivations or \ref{sec:Methods for computation of the YORP effect} for a summary.)

\begin{figure}
	\centering
	\includegraphics[width=0.48\textwidth]{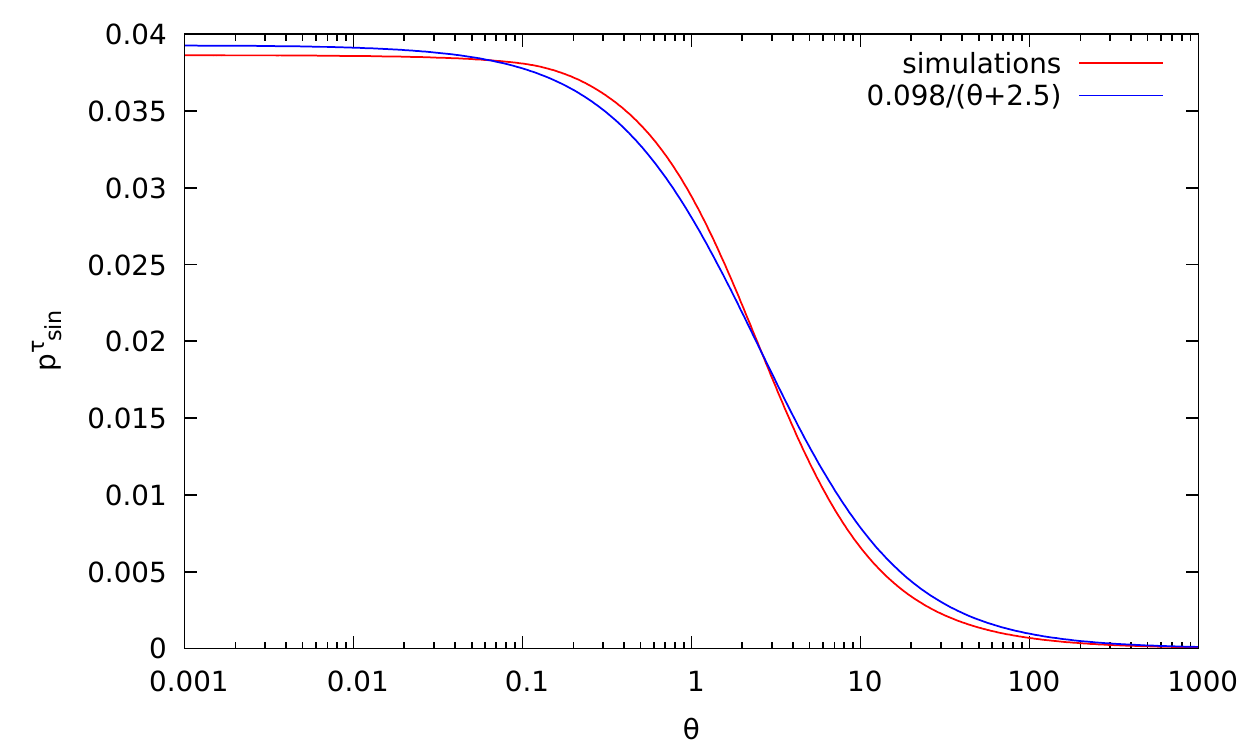}
	\includegraphics[width=0.48\textwidth]{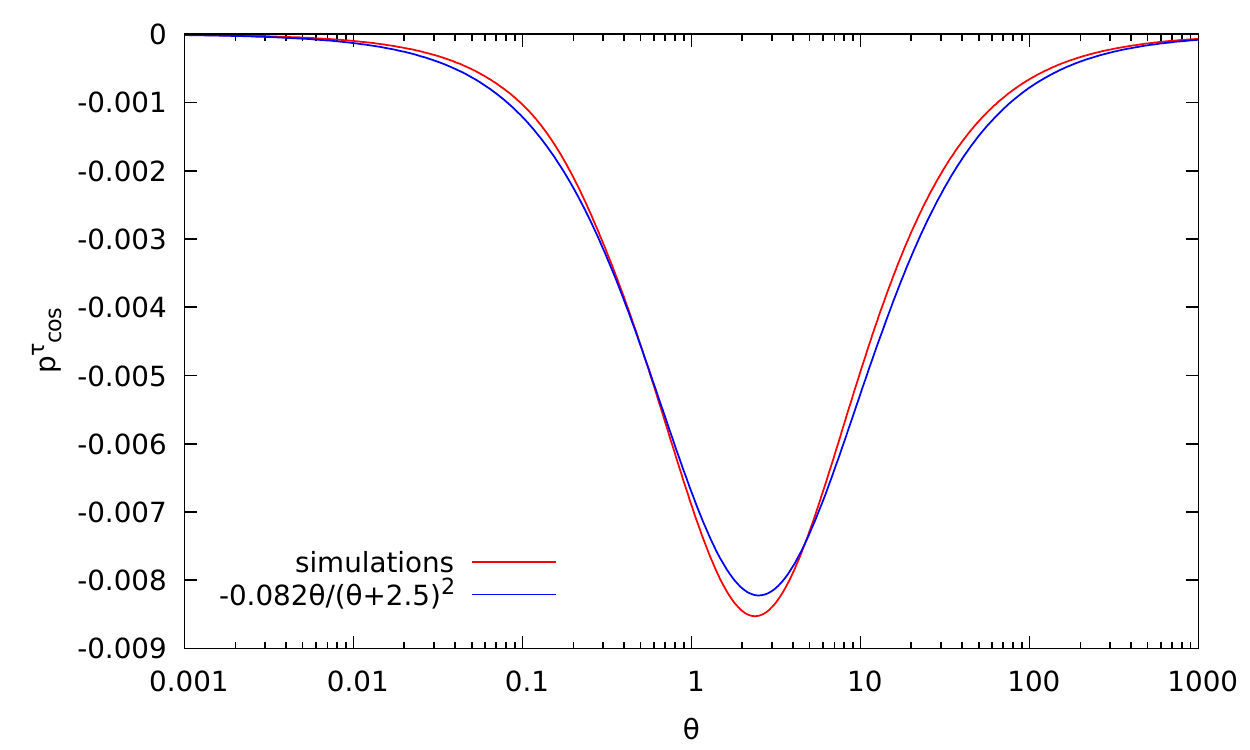}
	\caption{The numeric solution for the Fourier coefficients $p_\mathrm{sin}$ and $p_\mathrm{cos}$ of the dimensionless YORP pressures, and their fit by Eqn. (\ref{p_fit}).}
	\label{fig:p_theta}
\end{figure}

To compute $p^\tau_\mathrm{sin}(\psi, \varepsilon,\theta)$ and $p^\tau_\mathrm{cos}(\psi, \varepsilon,\theta)$, we use an algorithm similar to \cite{bbc10}.
We simulate a one-dimensional heat conductivity in a semi-space under the surface by decomposing the temperature into a Fourier series, expressing the boundary condition as a set of non-linear equations for the Fourier coefficients, and iteratively finding its solution.
Then the $p^\tau_\mathrm{sin}$ and $p^\tau_\mathrm{cos}$ are expressed in terms of the obtained Fourier series.
The results are in agreement with \cite{golubov16nyorp}, who evaluated the same functions, using finite difference method to solve the heat conductivity equation. Still, the method applied here works about three orders of magnitude faster, providing the same accuracy.

From our numeric simulations, it follows that a reasonable approximation to the functions $p^\tau_\mathrm{sin}$ and $p^\tau_\mathrm{cos}$ is given by the equations
\begin{eqnarray}
p^\tau_\mathrm{sin}\left(\psi, \varepsilon,\theta\right) = p_\mathrm{sin}(\theta)\sin 2\varepsilon\sin 2\psi,\nonumber\\
\label{p_t_approx}
p^\tau_\mathrm{cos}\left(\psi, \varepsilon,\theta\right) = p_\mathrm{cos}(\theta)\sin 2\varepsilon\sin 2\psi.
\end{eqnarray}
These equations basically express the principal non-vanishing terms of the Fourier decomposition of $p^\tau_\mathrm{sin}$ and $p^\tau_\mathrm{cos}$ as a function of $\psi$ and $\varepsilon$. 
Additional arguments for the validity of this approximation are given in \ref{sec:Methods for computation of the YORP effect}.
The numerically determined $p_\mathrm{sin}(\theta)$ and $p_\mathrm{cos}(\theta)$ are plotted in Figure \ref{fig:p_theta}.

From Section 3 of \cite{golubov16nyorp} we know the asymptotics of $p^\tau_\mathrm{sin}$ and $p^\tau_\mathrm{cos}$: $p_\mathrm{sin}\propto p_\mathrm{cos}\propto\theta^{-1}$ for $\theta\rightarrow\infty$, and
$p_\mathrm{sin}\propto\theta^0$,
$p_\mathrm{cos}\propto\theta^1$ for $\theta\rightarrow 0$.
We fit the numeric solution for $p^\tau_\mathrm{sin}$ and $p^\tau_\mathrm{cos}$ by the analytic expressions that have the correct asymptotic behavior (Figure \ref{fig:p_theta})
\begin{eqnarray}
p_\mathrm{sin}(\theta)=\frac{\tilde{p}_\mathrm{s}}{\theta+\theta_0},\nonumber\\
p_\mathrm{cos}(\theta)=-\frac{\tilde{p}_\mathrm{c}\theta}{(\theta+\theta_0)^2}
\label{p_fit}
\end{eqnarray}
The best-fit coefficients are about $\tilde{p}_\mathrm{s}=0.098$, $\tilde{p}_\mathrm{c}=0.082$, and $\theta_0=2.5$. 

With the aid of Eqs. (\ref{p_t_approx}) and (\ref{p_fit}), the expression for the obliquity component of YORP transforms into
\begin{eqnarray}
\tau_{\varepsilon} = \sin{2\varepsilon}\big{(}AC_\mathrm{sin}p_\mathrm{sin}(0)+\nonumber\\
+(1-A)C_\mathrm{sin}p_\mathrm{sin}(\theta)+(1-A)C_\mathrm{cos}p_\mathrm{cos}(\theta)\big{)}.
\label{T_eps_fit}
\end{eqnarray}
The coefficients $C_\mathrm{sin}$ and $C_\mathrm{cos}$ are expressed as integrals over the asteroid surface (see \ref{sec:Analytic estimates for the YORP coefficient}). Therefore, all the information about the asteroid shape needed to compute the YORP evolution is contained in these two coefficients. The coefficient $C_\mathrm{sin}$ is proportional to $C_\varepsilon$ from \cite{golubov19} and differs from it by the factor $p_\mathrm{sin}(0)$, whereas $C_\mathrm{cos}$ is a new concept that was absent in zero thermal inertia case.

The first term in Eqn. (\ref{T_eps_fit}) proportional to the albedo $A$ expresses the contribution to YORP from the light scattered by the asteroid. Zero argument of $p_\mathrm{sin}$ arises from the immediacy of light scattering, which is equivalent to no thermal inertia. The following two terms correspond to the re-emitted light, and thus proportional to the absorption fraction $1-A$.

\subsection{Investigation of the YORP coefficients}

To compute the YORP coefficients $C_\mathrm{sin}$ and $C_\mathrm{cos}$ and study the asteroid evolution, we take a sample of 5716 photometric shape models from DAMIT\footnote{DAMIT, https://astro.troja.mff.cuni.cz/projects/damit/} \cite{durech10},
29 radar shape models\footnote{Asteroid Radar Research. Asteroid Shape Models https://echo.jpl.nasa.gov/asteroids/shapes/shapes.html}, and 4 \textit{in situ} models of asteroids Eros, Itokawa, Bennu, and Ryugu\footnote{PDS Small Bodies Node. Shape Models of Asteroids, Comets, and Satellites, https://sbn.psi.edu/pds/shape-models/}. If several photometric or radar shape models of the same asteroid were present in the database, we processed them all independently, whereas for the \textit{in situ} models we used the ones with 196608 facets. We assumed that the $z$-axis of the shape model was the rotation axis of the asteroid, which in some cases implied a non-principal axis rotation.

\begin{figure}
	\centering
	\includegraphics[width=0.48\textwidth]{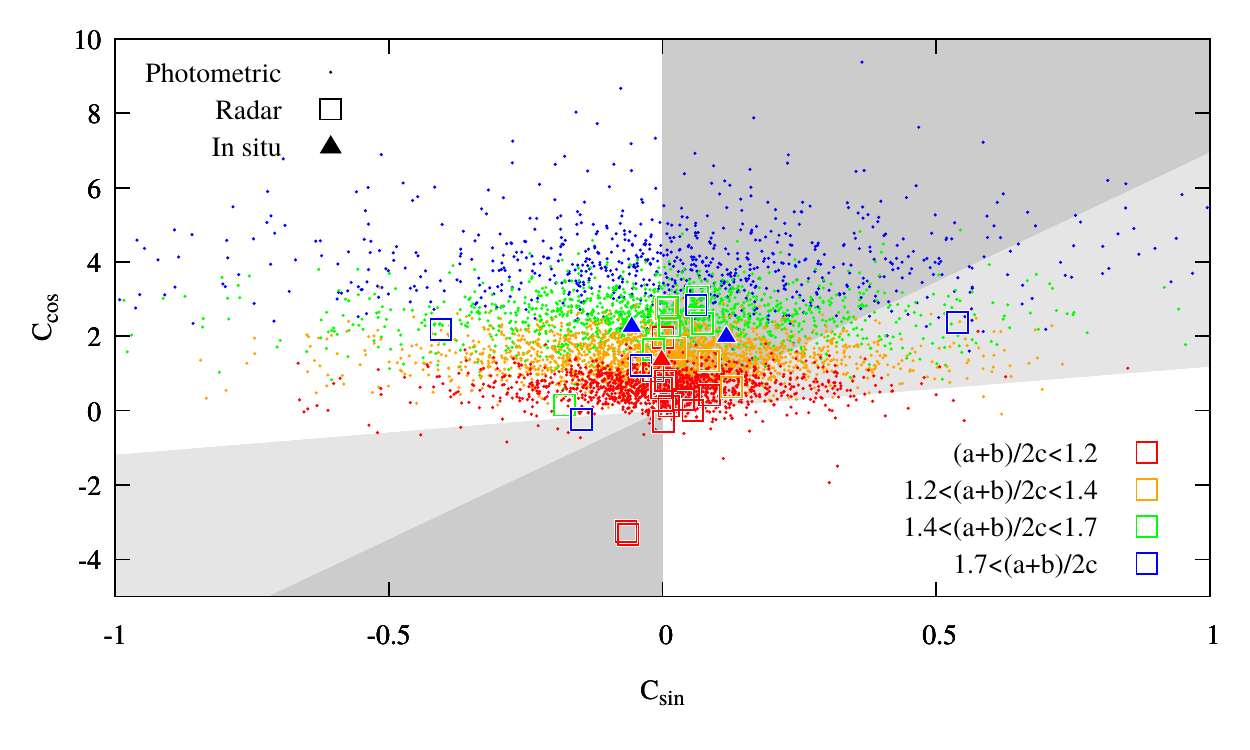}
	\caption{Comparison of $C_\mathrm{sin}$ and $C_\mathrm{cos}$ for different asteroids. Shading marks the areas where the asteroids can have rotational equilibria: for the albedo $A=0.5$ (dark gray) and $A=0$ (both dark and light gray). The points are color coded according to the pole flattening of asteroids $(a+b)/2c$, where  $a$, $b$ and $c$ denote the three semimajor axes of the asteroid, and rotation occurs around the $c$-axis.}
	\label{fig:sin_cos_ellipticity}
\end{figure}

Dependence between $C_\mathrm{sin}$ and $C_\mathrm{cos}$ is studied in Figure \ref{fig:sin_cos_ellipticity}. One can see that $C_\mathrm{cos}$ is positive in the predominant majority of cases, whereas $C_\mathrm{sin}$ has equal probabilities of being positive or negative.
The symmetric distribution of the points in the plot implies absence of correlation between $C_\mathrm{sin}$ and $C_\mathrm{cos}$. On the other hand, there is a strong correlation between the asteroid pole flattening and $C_\mathrm{cos}$, as revealed by the color coding. For example, the two overlapping red open squares that are the lowest points in the plot correspond to two models of asteroid 4179 Toutatis. This asteroid experiences tumbling, and the $z$-axis of its shape model is oriented in such unnatural way that $(a+b)/2c\approx 0.4$ is much less than unity, once again confirming the mentioned correlation. 

\begin{figure}
	\centering
	\includegraphics[width=0.48\textwidth]{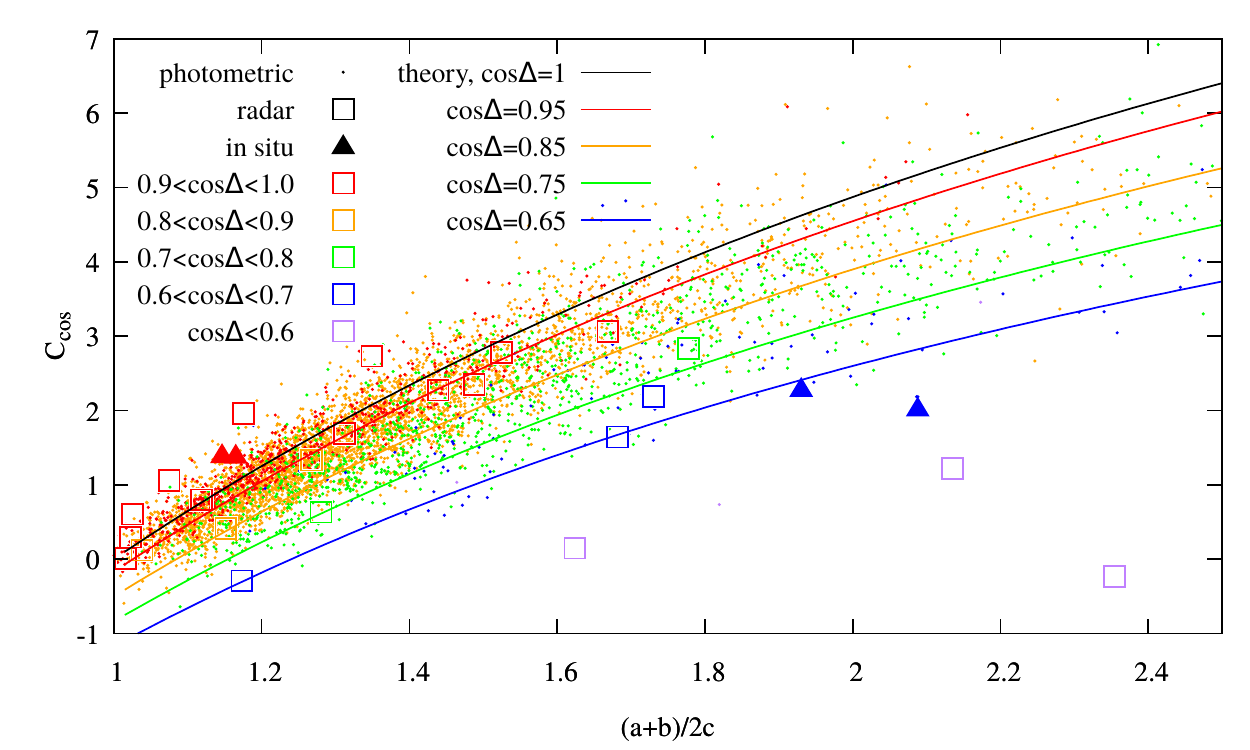}
	\caption{Dependence of $C_\mathrm{cos}$ on ellipticity. The color encodes the value of $\cos\Delta$, which serves as a proxy for the asteroid roughness. The simple theory for a smooth asteroid (black line) and its corrections for different values of $\Delta$ (colored lines) are explained in \ref{sec:Analytic estimates for the YORP coefficient}.}
	\label{fig:cos_ellipticity}
\end{figure}

To further study the dependence of $C_\mathrm{cos}$ on the pole flattening, we analyze Figure \ref{fig:cos_ellipticity}. There is a clear monotonous increase of $C_\mathrm{cos}$ with $(a+b)/2c$, which is analytically described within \ref{sec:Analytic estimates for the YORP coefficient}. The corresponding analytic formula is plotted by the black line. The agreement between this analytical line and numerically computed points can be further improved by accounting for the roughness of the asteroid surface. We characterize the roughness by the angle $\bar{\Delta}$, and color code the numeric points according to its value. The inclusion of $\bar{\Delta}$ provides a correction to the theory, which is also explained in \ref{sec:Analytic estimates for the YORP coefficient} and plotted in Figure \ref{fig:cos_ellipticity} by solid lines with the same color coding. Each line neatly crosses the cloud of points of the same color. The three purple squares in the lower right portion of the plot are a radar shape model of asteroid (8567) 1996 HW$_1$ and two different models of (216) Kleopatra. Both these asteroids have shapes of contact binaries, and correspondingly high angles $\bar{\Delta}$.

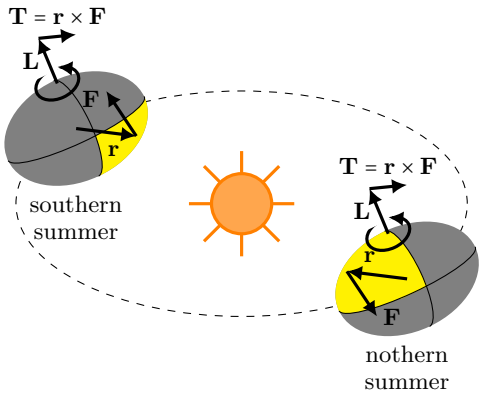
\begin{figure}
	\centering
	\begin{tikzpicture}
	\draw[orange,very thick] (-0.7,0)--(0.7,0);
	\draw[orange,very thick] (0,-0.7)--(0,0.7);
	\draw[orange,very thick] (-0.5,0.5)--(0.5,-0.5);
	\draw[orange,very thick] (0.5,0.5)--(-0.5,-0.5);
	\draw[dashed] (0,0) ellipse (3 and 1.5);
	\fill[color=orange!70] (0,0) circle [radius=0.4];
	\draw[orange,very thick] (0,0) circle [radius=0.4];
	\fill[rotate around={25:(-2.2,1)},gray] (-2.2,1) ellipse (1 and 0.7);
	\begin{scope}
	\path[clip] (-2.2,1) -- (-0.4,1.85) -- (-1.35,-0.8) -- cycle;
	\fill[rotate around={25:(-2.2,1)},yellow] (-2.2,1) ellipse (1 and 0.7);
	\end{scope}
	\fill[rotate around={25:(-2.2,1)},gray] (-2.2,1) ellipse (1 and 0.2);
	\fill[rotate around={25:(-2.2,1)},gray] (-2.2,1) ellipse (0.2 and 0.7);
	\begin{scope}
	\path[clip] (-4,0.15) -- (-0.4,1.85) -- (0,0) -- (-4,0) -- cycle;
	\draw[rotate around={25:(-2.2,1)}] (-2.2,1) ellipse (1 and 0.2);
	\end{scope}
	\begin{scope}
	\path[clip] (-3.05,2.8) -- (-1.35,-0.8) -- (0,-0.8) -- (0,2.8) -- cycle;
	\draw[rotate around={25:(-2.2,1)}] (-2.2,1) ellipse (0.2 and 0.7);
	\end{scope}
	\begin{scope}
	\path[clip] (-0.4,2.75) -- (-4,1.05) -- (-4.4,0) -- (-0.4,0) -- cycle;
	\draw[rotate around={25:(-2.45,1.6)},very thick] (-2.45,1.6) ellipse (0.3 and 0.2);
	\end{scope}
	\draw[-{Latex[length=2mm, width=2mm]},very thick] (-2.4,1.815)--(-2.45,1.82);
	\draw[-{Latex[length=2mm, width=2mm]},very thick] (-2.45, 1.6)--(-2.7,2.2);
	\draw[-{Latex[length=2mm, width=2mm]},very thick] (-2.7,2.2)--(-2.2,2.25);
	\draw[-{Latex[length=2mm, width=2mm]},very thick] (-2.2,1)--(-1.4,0.9);
	\draw[-{Latex[length=2mm, width=2mm]},very thick] (-1.4,0.9)--(-1.8,1.5);
	\node[] at (-2.8,1.9) {$\textbf{L}$};
	\node[] at (-2.45,2.5) {$\textbf{T}=\textbf{r}\times\textbf{F}$};
	\node[] at (-1.7,0.7) {$\textbf{r}$};
	\node[] at (-2,1.4) {$\textbf{F}$};
	\node[] at (-2.2,0) {southern};
	\node[] at (-2.2,-0.4) {summer};
	\fill[rotate around={25:(2.2,-1)},gray] (2.2,-1) ellipse (1 and 0.7);
	\begin{scope}
	\path[clip] (2.2,-1) -- (0.4,-1.85) -- (1.35,0.8) -- cycle;
	\fill[rotate around={25:(2.2,-1)},yellow] (2.2,-1) ellipse (1 and 0.7);
	\end{scope}
	\begin{scope}
	\path[clip] (2.47,-1.1) -- (0.67,-1.95) -- (1.62,0.7) -- cycle;
	\fill[rotate around={25:(2.2,-1)},yellow] (2.2,-1) ellipse (1 and 0.2);
	\fill[rotate around={25:(2.2,-1)},yellow] (2.2,-1) ellipse (0.2 and 0.7);
	\end{scope}
	\begin{scope}
	\path[clip] (4,-0.15) -- (0.4,-1.85) -- (0,-2) -- (4,-2) -- cycle;
	\draw[rotate around={25:(2.2,-1)}] (2.2,-1) ellipse (1 and 0.2);
	\end{scope}
	\begin{scope}
	\path[clip] (3.05,-2.8) -- (1.35,0.8) -- (5,0.8) -- (5,-2) -- cycle;
	\draw[rotate around={25:(2.2,-1)}] (2.2,-1) ellipse (0.2 and 0.7);
	\end{scope}
	\begin{scope}
	\path[clip] (4,0.75) -- (0.4,-0.95) -- (0,-2) -- (4,-2) -- cycle;
	\draw[rotate around={25:(1.95,-0.4)},very thick] (1.95,-0.4) ellipse (0.3 and 0.2);
	\end{scope}
	\draw[-{Latex[length=2mm, width=2mm]},very thick] (2.00,-0.185)--(1.95,-0.18);
	\draw[-{Latex[length=2mm, width=2mm]},very thick] (1.95,-0.4)--(1.7,0.2);
	\draw[-{Latex[length=2mm, width=2mm]},very thick] (1.7,0.2)--(2.2,0.25);
	\draw[-{Latex[length=2mm, width=2mm]},very thick] (2.2,-1)--(1.4,-0.9);
	\draw[-{Latex[length=2mm, width=2mm]},very thick] (1.4,-0.9)--(1.8,-1.5);
	\node[] at (1.6,-0.1) {$\textbf{L}$};
	\node[] at (1.95,0.5) {$\textbf{T}=\textbf{r}\times\textbf{F}$};
	\node[] at (1.7,-0.7) {$\textbf{r}$};
	\node[] at (2,-1.5) {$\textbf{F}$};
	\node[] at (2.2,-2) {nothern};
	\node[] at (2.2,-2.4) {summer};
	\end{tikzpicture}
	\caption{Physical explanation of the negative $C_\mathrm{cos}$ coefficient. The warmest part of the asteroid is the evening side of its summer hemisphere (colored in yellow). By computing the vector product of the radius-vector \textbf{r} directed into this area and the corresponding recoil light pressure force, we get the torque \textbf{T}, which always decreases the absolute value of the asteroid's obliquity.}
	\label{fig:Csos-psysical-explanation}
\end{figure}

The predominance of positive values of $C_\mathrm{cos}$ has a simple physical explanation, illustrated in Figure \ref{fig:Csos-psysical-explanation}. Consider a flattened ellipsoidal asteroid. The highest temperature is attained on the evening side of the summer hemisphere.
Therefore, this is the side of the asteroid that experiences the highest recoil light pressure.
Using the right-hand rule, one can check that for both the southern summer and the northern summer this force creates a negative torque, which decreases the obliquity. 
According to Eqn. (\ref{dimensionless_evolution_equation2}), this corresponds to $\tau_\varepsilon<0$. 
Then one can look at Eqn. (\ref{T_eps_fit}), note that $C_\mathrm{sin}=0$ for ellipsoidal asteroid (see \ref{sec:Analytic estimates for the YORP coefficient} for the proof), and conclude that $C_\mathrm{cos}p_\mathrm{cos}<0$.
As $p_\mathrm{cos}$ is always negative (see Figure \ref{fig:p_theta}), $C_\mathrm{cos}$ has to be positive, just as it can be seen in Figure \ref{fig:sin_cos_ellipticity}. 

Moreover, the YORP torque is zero for spherical asteroids, where the light pressure forces have zero lever arm, and it rises for more flattened asteroids as the lever arm increases, which agrees with the monotonic growth of $C_\mathrm{cos}$ as a function of flattening in Figure \ref{fig:cos_ellipticity}.
Naturally, this effect also vanishes for very fast and very slow rotators, as they do not have a significant temperature differences between the evening and the morning sides. This agrees with the asymptotic behavior of $p_\mathrm{cos}$, which vanishes in the limits $\theta\rightarrow 0$ and $\theta\rightarrow\infty$.

It is important to note, that $C_\mathrm{cos}$ is non-zero even for a perfect ellipsoid, while $C_\mathrm{sin}$ is only produced by its asymmetry, which is usually slight.
It can explain why in most of the cases $C_\mathrm{cos}$ is about an order of magnitude bigger than $C_\mathrm{sin}$.

\section{YORP evolution and eqilibria}
\label{sec:YORP eqilibria}
\subsection{Overall YORP evolution}
From \cite{golubov19} we know an approximate expression for $\tau_z$, which in our present notations looks like
\begin{equation}
\tau_z=\frac{C_\mathrm{sin}\tilde{p}_\mathrm{s}}{\alpha\theta_0}\left(\cos{2\varepsilon}+\beta\right),
\label{T_z_fit}
\end{equation}
where the coefficients $\alpha\approx 0.72$ and $\beta\approx 0.33$.
When Eqs. (\ref{T_eps_fit}) and (\ref{T_z_fit}) are substituted into Eqs. (\ref{dimensionless_evolution_equation1}) and (\ref{dimensionless_evolution_equation2}), a full set of evolutionary equations is obtained. It describes $\theta$ and $\varepsilon$ as functions of time $t$.

In this section, we will investigate the typical solutions of these equations. As $C_\mathrm{cos}>0$ in the majority of cases (see Figure \ref{fig:sin_cos_ellipticity}), this is what we assume henceforth.
On the other hand, the sign of $C_\mathrm{sin}$ seems to be positive and negative with equal probabilities, thus we consider both cases.

In the case $C_\mathrm{sin}<0$, the topology of the solution is the same as in \cite{golubov19}.
The asteroids start at small rotation rates, accelerate their rotation, and then slow it down, if not disrupted by the centrifugal forces on the way.
The most important quantitative differences from \cite{golubov19} occur at $\theta\sim 1$, where the large $C_\mathrm{cos}$ term causes obliquity to evolve much faster than the rotation rate.

\begin{figure}
	\centering
	\begin{tikzpicture}
	\draw[-{Latex[length=2mm, width=2mm]},very thick] (0,0)--(0,7);
	\draw[-{Latex[length=2mm, width=2mm]},very thick] (0,0)--(7,0);
	\draw[-{Latex[length=2mm, width=2mm]},very thick,blue] (1.35,0.5)--(2.35,0.5);
	\draw[-{Latex[length=2mm, width=2mm]},very thick,green] (1.35,0.5)--(1.35,1.5);
	\draw[-{Latex[length=2mm, width=2mm]},very thick,red] (1.35,0.5)--(2.35,1.5);
	\draw[-{Latex[length=2mm, width=2mm]},very thick,blue] (2.35,3)--(1.35,3);
	\draw[-{Latex[length=2mm, width=2mm]},very thick,green] (1.35,3)--(1.35,4);
	\draw[-{Latex[length=2mm, width=2mm]},very thick,red] (2.35,3)--(1.35,4);
	\draw[-{Latex[length=2mm, width=2mm]},very thick,blue] (1.35,5.5)--(2.35,5.5);
	\draw[-{Latex[length=2mm, width=2mm]},very thick,green] (1.35,5.5)--(1.35,6.5);
	\draw[-{Latex[length=2mm, width=2mm]},very thick,red] (1.35,5.5)--(2.35,6.5);
	\draw[-{Latex[length=2mm, width=2mm]},very thick,blue] (4.35,0.5)--(5.35,0.5);
	\draw[-{Latex[length=2mm, width=2mm]},very thick,green] (4.35,1.5)--(4.35,0.5);
	\draw[-{Latex[length=2mm, width=2mm]},very thick,red] (4.35,1.5)--(5.35,0.5);
	\draw[-{Latex[length=2mm, width=2mm]},very thick,blue] (5.35,3)--(4.35,3);
	\draw[-{Latex[length=2mm, width=2mm]},very thick,green] (4.35,4)--(4.35,3);
	\draw[-{Latex[length=2mm, width=2mm]},very thick,red] (5.35,4)--(4.35,3);
	\draw[-{Latex[length=2mm, width=2mm]},very thick,blue] (4.35,5.5)--(5.35,5.5);
	\draw[-{Latex[length=2mm, width=2mm]},very thick,green] (4.35,6.5)--(4.35,5.5);
	\draw[-{Latex[length=2mm, width=2mm]},very thick,red] (4.35,6.5)--(5.35,5.5);	
	\draw[] (6,0)--(6,7);
	\draw[green,dashed] (3.7,0)--(3.7,6.7);
	\draw[blue,dashed] (0,2)--(6,2);
	\draw[blue,dashed] (0,5)--(6,5);
	\node[blue] at (6.7,2) {$\tau_\varepsilon=0$};
	\node[blue] at (6.7,5) {$\tau_\varepsilon=0$};
	\node[green] at (3.7,6.9) {$\tau_z=0$};
	\node[] at (6.9,0.3) {$\varepsilon$};
	\node[] at (-0.25,6.9) {$\theta$};
	\node[] at (-0.25,5) {$\theta_0^{+}$};
	\node[] at (-0.25,2) {$\theta_0^{-}$};
	\node[] at (3.75,-0.25) {$\varepsilon_0$};
	\node[] at (6,-0.25) {$90^\circ$};
	\node[] at (-0.1,-0.2) {$0$};
	\fill[color=red] (3.7,5) circle [radius=0.1];
	\draw[color=red] (3.7,5) circle [radius=0.1];
	\draw[-{Latex[length=2mm, width=2mm]},very thick,red] (3.3,5.1)--(3.6,5.4);	
	\draw[-{Latex[length=2mm, width=2mm]},very thick,red] (3.8,5.4)--(4.1,5.1);	
	\draw[-{Latex[length=2mm, width=2mm]},very thick,red] (4.1,4.9)--(3.8,4.6);	
	\draw[-{Latex[length=2mm, width=2mm]},very thick,red] (3.6,4.6)--(3.3,4.9);
	\fill[white,very thick] (3.7,2) circle [radius=0.1];
	\draw[red,very thick] (3.7,2) circle [radius=0.1];
	\draw[-{Latex[length=2mm, width=2mm]},very thick,red] (3.6,2.1)--(3.3,2.4);	
	\draw[-{Latex[length=2mm, width=2mm]},very thick,red] (4.1,2.4)--(3.8,2.1);	
	\draw[-{Latex[length=2mm, width=2mm]},very thick,red] (3.8,1.9)--(4.1,1.6);	
	\draw[-{Latex[length=2mm, width=2mm]},very thick,red] (3.3,1.6)--(3.6,1.9);	
	\end{tikzpicture}
	\caption{A sketch of the typical asteroid evolution in the case $C_\mathrm{sin}>0$. Green arrows represent the evolutionary direction for the thermal parameter $\theta$, blue arrows -- for the obliquity $\varepsilon$, red arrows -- directions of the overall asteroid evolution. Dashed lies separate regions with different signs of $\tau_z$ and $\tau_\varepsilon$. Open red circle marks a saddle point, and filled red circles marks a focus on the evolution diagram.}
	\label{fig:evolution}
\end{figure}
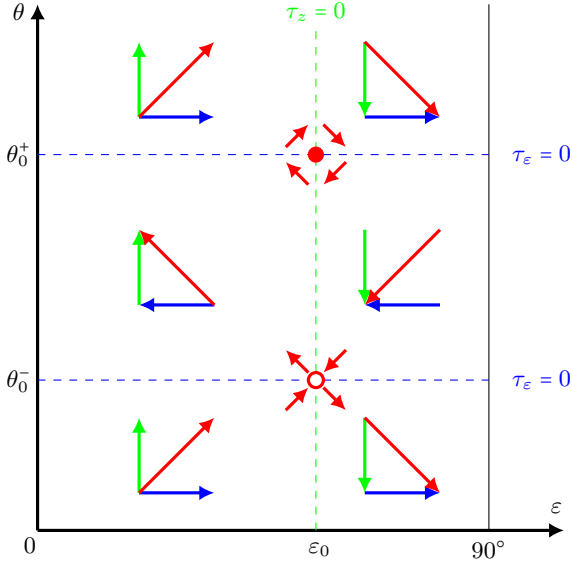

In the opposite case of $C_\mathrm{sin}>0$, a qualitatively new behavior can arise, as it is shown in Figure \ref{fig:evolution}.
At $\theta\ll 1$ and $\theta\gg 1$, the contribution from $C_\mathrm{sin}$ dominates in Eqn. (\ref{T_eps_fit}), causing $\varepsilon$ to increase. On the other hand, at $\theta\sim 1$, the major contribution to Eqn. (\ref{T_eps_fit}) can arise from $C_\mathrm{cos}$, leading to $\dot{\varepsilon}<0$. These areas with different signs of $\tau_\varepsilon$ are separated by the curves where $\tau_\varepsilon=0$, which in our approximation are straight. They are shown in Figure \ref{fig:evolution} with blue dashed lines. Additionally, Eqn. (\ref{T_z_fit}) turns into zero at $\varepsilon\approx 55^\circ$, which is shown in the figure with a green dashed line. These lines split the phase plane into parts where $\theta$ and $\varepsilon$ change in different directions, shown with colored arrows in Figure \ref{fig:evolution}.
On the boundaries, two equilibrium points originate. The lower one, at $(\theta_0^{-},\varepsilon_0)$, is an unstable saddle point. The upper equilibrium, at $(\theta_0^{+},\varepsilon_0)$, is a focal point, whose stability requires additional investigation. We postpone the discussion of their stability till Subsection 3.3, and first discuss under which circumstances and in which rotation states such equilibria occur.

\subsection{Existence of YORP equilibria}

To find the equilibria, we substitute Eqn. (\ref{p_fit}) into Eqn. (\ref{T_eps_fit}), equate its right-hand side to zero, and solve for $\theta$. The roots of the resulting quadratic equation are
\begin{eqnarray}
\hspace{-0.8cm}
\theta^{\pm}_{0}=\frac{\theta_0}{2A\tilde{p}_\mathrm{s}C_\mathrm{sin}}\Big{(}\tilde{p}_\mathrm{c}(1-A)C_\mathrm{cos}-\tilde{p}_\mathrm{s}(1+A)C_\mathrm{sin}\pm\nonumber\\
\hspace{-0.8cm}
\pm \sqrt{-4\tilde{p}_\mathrm{s}^2AC_\mathrm{sin}^2+(\tilde{p}_\mathrm{c}(1-A)C_\mathrm{cos}-\tilde{p}_\mathrm{s}(1+A)C_\mathrm{sin})^2}\Big{)}.
\label{roots_cos}
\end{eqnarray}
The equilibria exist when the expression under the square root is positive, which is equivalent to the following condition,
\begin{equation}
\frac{C_\mathrm{cos}}{C_\mathrm{sin}}\geq\frac{1+\sqrt{A}}{1-\sqrt{A}}\frac{\tilde{p}_\mathrm{s}}{\tilde{p}_\mathrm{c}}.
\label{equilibria_condition}
\end{equation}
The areas where this condition is met are shown in shades of gray in Figure \ref{fig:sin_cos_ellipticity}.
We see that the majority of points with $C_\mathrm{sin}>0$ satisfy this condition. This result comes naturally from the fact that for most of the asteroids $C_\mathrm{cos}\gg|C_\mathrm{sin}|$ (see Figure \ref{fig:sin_cos_ellipticity}).

\begin{figure}
	\centering
	\includegraphics[width=0.48\textwidth]{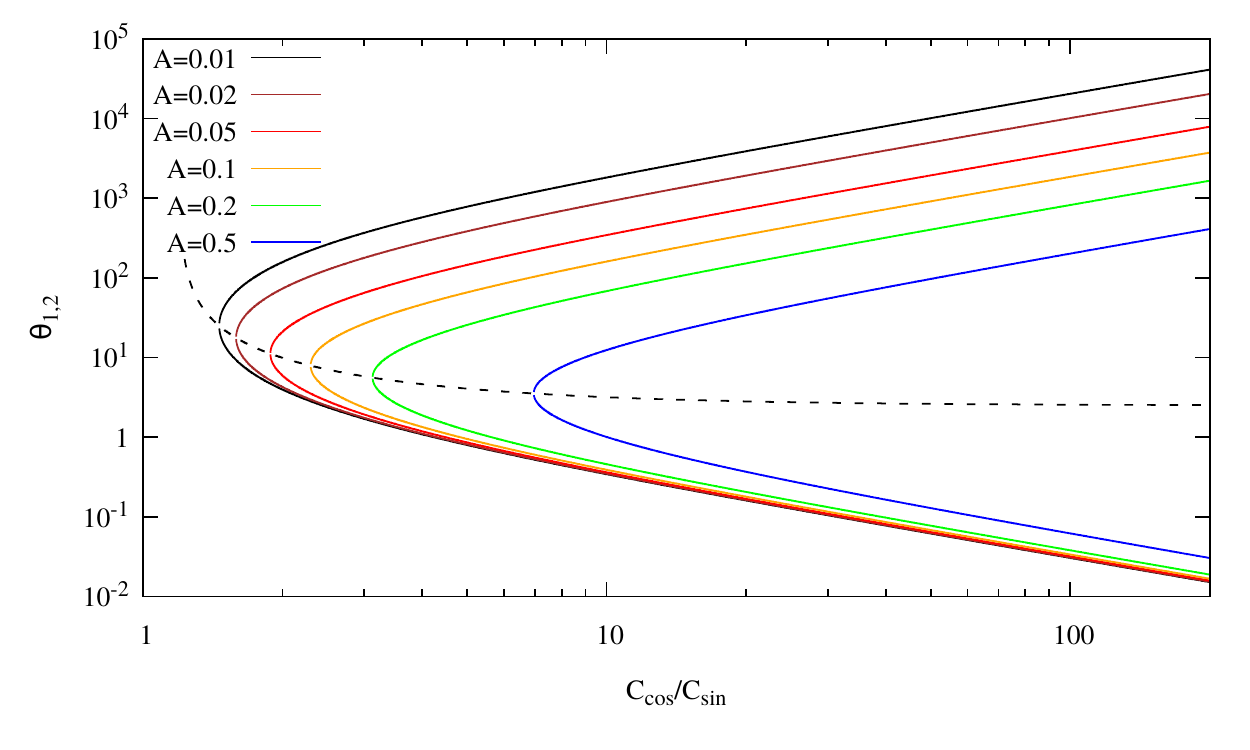}
	\caption{Equilibrium points as given by by Eqn. (\ref{roots_cos}). Albedo is color-coded. The dashed black line marks the positions of bifurcation points.}
	\label{fig:roots_cos}
\end{figure}

Eqn. (\ref{roots_cos}) is illustrated in Figure \ref{fig:roots_cos}, which shows the equilibrium points $\theta_1$ and $\theta_2$ as functions of $C_\mathrm{cos}/C_\mathrm{sin}$ for several fixed values of albedo $A$.
As the ratio $C_\mathrm{cos}/C_\mathrm{sin}$ increases, a bifurcation occurs, in which two equilibria originate. 

It is easy to understand why YORP equilibria always appear in pairs. In the absence of tangential YORP $\tau_z$ is independent of $\theta$. Thus the condition $\tau_z=0$ prescribes the value of $\varepsilon$. With $\varepsilon$ being fixed, the second condition for the equilibrium $\tau_\varepsilon=0$ turns into equation only for $\theta$.
Applying the asymptotics for $p_\mathrm{sin}$ and $p_\mathrm{cos}$ to Eqn. (\ref{T_eps_fit}), one can see that $\tau_{\varepsilon}|_{\theta=\infty}/\tau_{\varepsilon}|_{\theta=0} = A$.
Therefore, $\tau_\varepsilon$ has the same sign at $0$ and $\infty$.
Due to the continuity of the function $\tau_\varepsilon(\theta)$, it implies that the total number of sign reversals (i.e. equilibria) is even.
This fact was omitted by \cite{capek04}, who unphysically assumed $A=0$, and thus for many asteroids observed only one sign reversal. 

\begin{figure}
	\centering
	\begin{tikzpicture}	
	\begin{scope}
	\path[clip] (0,0) -- (0,6) -- (7,6) -- (7,0) -- cycle;
	\shade[shading=radial, inner color=orange] (2.2,3) rectangle (3,5);
	\shade[shading=radial, inner color=purple] (5,3) rectangle (3.5,5);
	\draw[very thick, dashed,red] (-1.5,-100)--(8.5,-100);
	\draw[very thick, dashed,red] (-1.5,-100)--(8.5,-100);
	\draw[very thick, dashed,green] (-1.5,4.30512)--(8.5,-7.42875);
	\draw[-{Latex[length=2mm, width=2mm]},very thick,green] (1.15,1.2)--(0.85,0.95);
	\draw[very thick, dashed,green] (-1.5,-100)--(8.5,-100);
	\draw[very thick, dashed,blue] (-1.5,7.39683)--(8.5,-4.33705);
	\draw[-{Latex[length=2mm, width=2mm]},very thick,blue] (3.45,1.55)--(3.15,1.3);
	\draw[very thick, dashed,blue] (-1.5,-100)--(8.5,-100);
	\draw[very thick,red] (-1.5,-100)--(8.5,-100);
	\draw[very thick,red] (-1.5,-100)--(8.5,-100);
	\draw[very thick,green] (-1.5,9.52018)--(8.5,-2.2137);
	\draw[very thick,dotted,green] (0.55,1.75)--(2.93,3.73);
	\draw[-{Latex[length=2mm, width=2mm]},very thick,green] (3.23,3.98)--(2.93,3.73);
	\draw[very thick,green] (-1.5,3.54372)--(8.5,-8.19016);
	\draw[-{Latex[length=2mm, width=2mm]},very thick,green] (0.25,1.5)--(0.55,1.75);
	\draw[very thick,blue] (-1.5,12.6119)--(8.5,0.878009);
	\draw[-{Latex[length=2mm, width=2mm]},very thick,blue] (6.45,3.25)--(6.15,3);
	\draw[very thick,blue] (-1.5,0.45201)--(8.5,-11.2819);
	\draw[dashed,red] (-1.5,1.20981)--(8.5,-10.5241);
	\draw[dashed,red] (-1.5,-100)--(8.5,-100);
	\draw[dashed,green] (-1.5,4.44301)--(8.5,-7.29086);
	\draw[-{Latex[length=2mm, width=2mm]},green] (1.55,0.87)--(1.25,0.62);
	\draw[dashed,green] (-1.5,-100)--(8.5,-100);
	\draw[dashed,blue] (-1.5,7.46821)--(8.5,-4.26567);
	\draw[-{Latex[length=2mm, width=2mm]},blue] (3.75,1.3)--(3.45,1.05);
	\draw[dashed,blue] (-1.5,-100)--(8.5,-100);
	\draw[red] (-1.5,6.42486)--(8.5,-5.30901);
	\draw[-{Latex[length=2mm, width=2mm]},red] (3.02,1.13)--(2.72,0.88);
	\draw[red] (-1.5,4.41491)--(8.5,-7.31897);
	\draw[-{Latex[length=2mm, width=2mm]},red] (2,0.3)--(2.3,0.55);
	\draw[very thick,dotted,red] (2.3,0.55)--(2.72,0.88);
	\draw[green] (-1.5,9.65807)--(8.5,-2.07581);
	\draw[-{Latex[length=2mm, width=2mm]},green] (3.57,3.7)--(3.27,3.45);
	\draw[green] (-1.5,1.1817)--(8.5,-10.5522);
	\draw[blue] (-1.5,12.6833)--(8.5,0.949389);
	\draw[-{Latex[length=2mm, width=2mm]},blue] (6.75,3)--(6.45,2.75);
	\draw[blue] (-1.5,-100)--(8.5,-100);
	\end{scope}
	\draw[-{Latex[length=2mm, width=2mm]},very thick] (0,0)--(0,6);
	\draw[-{Latex[length=2mm, width=2mm]},very thick] (0,0)--(7,0);
	\draw[very thick] (0.4,0.1)--(0.4,-0.1);
	\draw[very thick] (2.5,0.1)--(2.5,-0.1);
	\draw[very thick] (4.4,0.1)--(4.4,-0.1);
	\draw[very thick] (6.5,0.1)--(6.5,-0.1);
	\node[] at (0.5,-0.3) {0.3};	
	\node[] at (2.5,-0.3) {1};	
	\node[] at (4.5,-0.3) {3};	
	\node[] at (6.5,-0.3) {10};	
	\draw[very thick] (-0.1,0.68)--(0.,0.68);	
	\draw[very thick] (-0.1,1.98)--(0.,1.98);	
	\draw[very thick] (-0.1,3.28)--(0.,3.28);	
	\draw[very thick] (-0.1,4.58)--(0.,4.58);
	\node[] at (-0.5,0.68) {$10^2$};	
	\node[] at (-0.5,1.98) {$10^3$};	
	\node[] at (-0.5,3.28) {$10^4$};	
	\node[] at (-0.5,4.58) {$10^5$};	
	\draw[very thick] (0,0.4)--(0.1,0.4);
	\draw[very thick] (0,2.7)--(0.1,2.7);
	\draw[very thick] (0,4.5)--(0.1,4.5);
	\draw[very thick] (0,5.6)--(0.1,5.6);
	\node[] at (0.5,0.4) {min};	
	\node[] at (0.5,2.7) {hour};	
	\node[] at (0.5,4.5) {day};	
	\node[] at (0.5,5.6) {week};
	\node[] at (6.9,0.3) {$a$, AU};	
	\node[] at (-0.45,5.9) {$P$, s};
	\draw[very thick,red] (-0.8,-0.7)--(0,-0.7);
	\draw[very thick,green] (-0.8,-1.2)--(0,-1.2);
	\draw[very thick,blue] (-0.8,-1.7)--(0,-1.7);	
	\node[] at (1.25,-0.7) {$C_\mathrm{cos}/C_\mathrm{sin}$=2.12};	
	\node[] at (1.25,-1.2) {$C_\mathrm{cos}/C_\mathrm{sin}$=13.2};	
	\node[] at (1.25,-1.7) {$C_\mathrm{cos}/C_\mathrm{sin}$=173};
	\draw[very thick,dashed] (3,-1.2)--(3.8,-1.2);
	\draw[very thick] (3,-1.7)--(3.8,-1.7);	
	\node[] at (3.1,-0.7) {$\Gamma$:};	
	\node[] at (4,-1.2) {3};	
	\node[] at (4.2,-1.7) {300};
	\draw[] (5,-1.2)--(5.8,-1.2);
	\draw[very thick] (5,-1.7)--(5.8,-1.7);	
	\node[] at (5.1,-0.7) {$A$:};	
	\node[] at (6.2,-1.2) {0.05};		
	\node[] at (6.2,-1.7) {0.25};	
	\node[color=orange] at (2.6,5.2) {NEAs};	
	\node[color=purple] at (4.25,5.2) {MBAs};	
	\end{tikzpicture}
	\caption{Equilibrium rotation periods $P$ corresponding to different heliocentric distances $a$. Different colors of lines correspond to different ratios of the YORP coefficients, which are determined by the asteroid shape. Type of the lines shows the value of thermal inertia $\Gamma$, and the thickness shows the albedo $A$. The arrows are directed towards the area where $\tau_\varepsilon$ is negative, and the dotted line crosses the area if both boundaries are seen in the plot. Shaded areas show typical ranges of $P$ and $a$ for near-Earth and main-belt asteroids.}
	\label{fig:scales}
\end{figure}
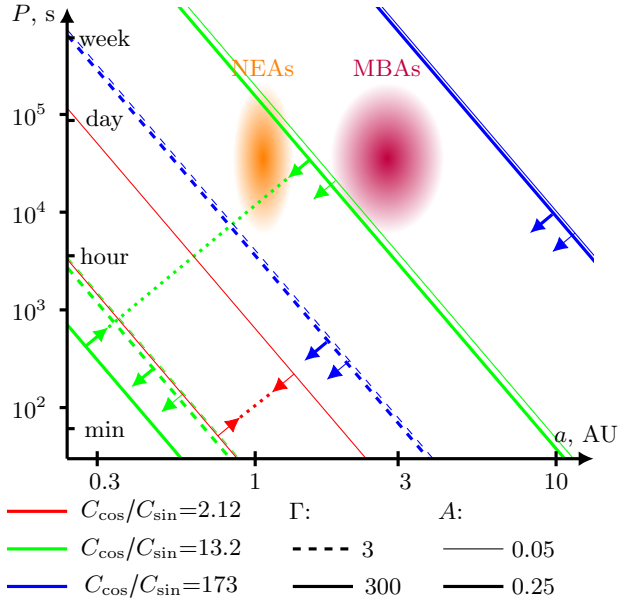

The rotation periods $P_{1,2}$ corresponding to the equilibrium thermal parameters $\theta_{1,2}$ are depicted in Figure \ref{fig:scales}. The horizontal axis shows different heliocentric distances. The three colors show different values of the ratio $C_\mathrm{cos}/C_\mathrm{sin}$, corresponding to the 10th, 50th, and 90th percentile of the distribution of studied DAMIT asteroids. Two thermal inertias $\Gamma$ are shown with different types of lines, corresponding to the lower and upper boundary of their typical values \citep{hanus18}. Two albedos $A=0.05$ and $A=0.25$ were chosen to represent low- and high-albedo asteroids \citep{albedo} and depicted with thin and thick lines correspondingly.

The plotted lines show where $\tau_\varepsilon$ changes sign, and small arrows point to where the sign is negative. To put it in a simple way, one can assume that in the direction of the arrows, $C_\mathrm{cos}$ dominates over $C_\mathrm{sin}$, and the result is the relatively fast alignment of the asteroid equatorial planes with their orbital planes, i.e. $\varepsilon=0^\circ$ or $180^\circ$. In the direction opposite to the arrows, $C_\mathrm{sin}$ dominates over $C_\mathrm{cos}$, resulting into similar probabilities of $\varepsilon=0^\circ/180^\circ$ and $\varepsilon=90^\circ$ and the evolution similar to the one described for the low-thermal-inertia limit by \cite{golubov19}.

One can see that depending on the values of parameters, the evolutionary regime can be very different.
For most of the high thermal inertia near-Earth asteroids $C_\mathrm{cos}$ is indeed dominant, whereas in the main asteroid belt this is true for only $\sim$50\% of the bodies. Among low thermal inertia asteroids $C_\mathrm{cos}$ dominates for $\sim$50\% of the near-Earth asteroids, but loses to $C_\mathrm{sin}$ for the overwhelming majority of the main belt asteroids.
We must conclude that the widely acknowledged alignment of asteroid equatorial and orbital planes due to YORP is only partially true and does not describe the entire asteroid population.

The lines with downward-pointing arrows correspond to potentially stable equilibria, similar in kind to the filled red circle in Figure \ref{fig:evolution}. Such equilibria are physically feasible only if they result into realistic rotation periods, hours to days, to avoid both tumbling or rotational disruption. 

Typical ranges of such periods for near-Earth asteroids (NEAs) and main-belt asteroids (MBAs) are marked in the plot. It can be seen from the plot that NEAs can have equilibria with realistic periods if they have high thermal inertias $\Gamma$ and the most probable $C_\mathrm{cos}/C_\mathrm{sin}$ ratios, or if they have low $\Gamma$ and high $C_\mathrm{cos}/C_\mathrm{sin}$. MBAs are expected to have realistic periods if they have high $\Gamma$ and beyond average $C_\mathrm{cos}/C_\mathrm{sin}$ ratios.

\begin{figure}
	\centering
	\includegraphics[width=0.45\textwidth]{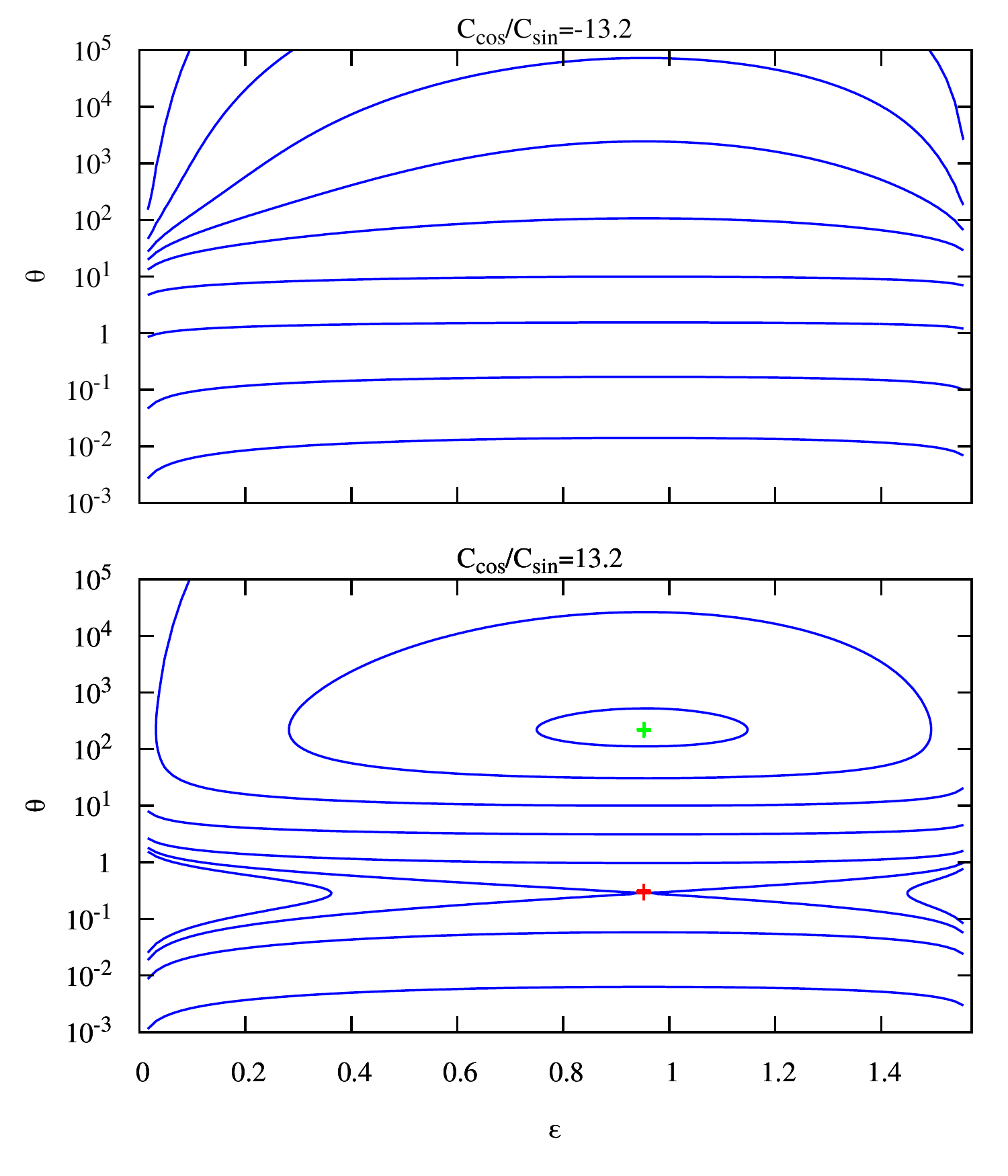}
	\caption{Examples of phase portraits of asteroids in the $\varepsilon$--$\theta$ space for two different values of $C_\mathrm{cos}/C_\mathrm{sin}$ and albedo $A=0.1$. Evolutionary tracks for asteroids as given by Eqn. (\ref{factorized_equations_solution}) are plotted with blue lines. The topology of phase curves in the upper panel is trivial, while the lower panel demonstrates a more interesting behavior with an unstable saddle point (marked with a red cross) and a central point (marked with a green cross).}
	\label{fig:obliquity_evolution}
\end{figure}

\subsection{Stability of YORP equilibria}

The physical significance of the focal equilibrium point depends on its stability. Unfortunately for our analytic model, it does not allow to conclude whether the focus is stable or not. The full set of evolutionary equations is obtained by substituting Eqs. (\ref{T_z_fit}) and (\ref{T_eps_fit}) into Eqs. (\ref{dimensionless_evolution_equation1}) and (\ref{dimensionless_evolution_equation2}) correspondingly:
\begin{eqnarray}
\frac{d\theta^2}{dt} = \frac{1}{i_z}\frac{C_\mathrm{sin}\tilde{p}_\mathrm{s}}{\alpha\theta_0}\left(\cos(2\varepsilon)+\beta\right),\nonumber\\
\frac{d\varepsilon}{dt} = \frac{1}{\theta^2}\frac{1}{i_z}\sin(2\varepsilon)\big{(}A C_\mathrm{sin}p_\mathrm{sin}(0)+\nonumber\\
+(1-A) C_\mathrm{sin}p_\mathrm{sin}(\theta)+(1-A) C_\mathrm{cos}p_\mathrm{cos}(\theta)\big{)}.
\label{factorized_equations}
\end{eqnarray}
The right-hand sides of the equations are factorized. They are products of functions depending on either $\varepsilon$ or $\theta$. We divide the first equation by the second one and separate the variables. The integral of the resulting expression gives an implicit solution of the system Eqs. (\ref{factorized_equations}),
\begin{eqnarray}
\frac{1}{4\alpha}(\beta\ln\tan\varepsilon+\ln\sin{2\varepsilon})-\ln\theta +\nonumber\\
+ (1-A)\ln(\theta+\theta_0) - \frac{(1-A)C_\mathrm{cos}\tilde{p}_\mathrm{c} \theta_0}{C_\mathrm{sin}\tilde{p}_\mathrm{s}(\theta+\theta_0)}= Const.
\label{factorized_equations_solution}
\end{eqnarray}

This equation provides an implicit solution of Eqs. (\ref{factorized_equations}), which is plotted in Figure \ref{fig:obliquity_evolution}. The two panels show two values of $C_\mathrm{cos}/C_\mathrm{sin}$, negative and positive, with the absolute value at the median of distribution of the DAMIT shapes. 

The lower parts of both panels ($\theta\ll 1$) show the geometry of phase trajectories described by \cite{golubov19}. The upper parts of the two panels is essentially the same, but with the factor of $1/A$ slower evolution in obliquity. 

In the middle part of the upper panel ($\theta\sim 1$), the evolution is also similar to \cite{golubov19} but even faster due to the contribution from the $C_\mathrm{cos}$ term. Therefore, for negative $C_\mathrm{cos}/C_\mathrm{sin}$, the entire phase portrait has a trivial topology. 

On the other hand, in the middle of the lower panel ($C_\mathrm{cos}/C_\mathrm{sin}>0$) two equilibrium points can appear, with a complex geometry of phase trajectories around them. The lower equilibrium is an unstable saddle point. The trajectories around the upper equilibrium are closed in our model. Thus it is neither a stable focus, nor an unstable focus, but a neutral center. This result though is model-dependent and breaks if higher-order terms are taken into account.

Therefore, the factorization of Eqs. (\ref{factorized_equations}) creates both an opportunity for analytic solution and an obstacle for stability analysis. 
In a more realistic theory, more Fourier terms should be taken in Eqn. (\ref{T_eps_fit}), thus breaking the factorization. This analytic approach will be further explored in our future article, while now we limit ourselves to a simple illustration of stability in one individual case.

\begin{figure}
	\centering
	\includegraphics[width=0.48\textwidth]{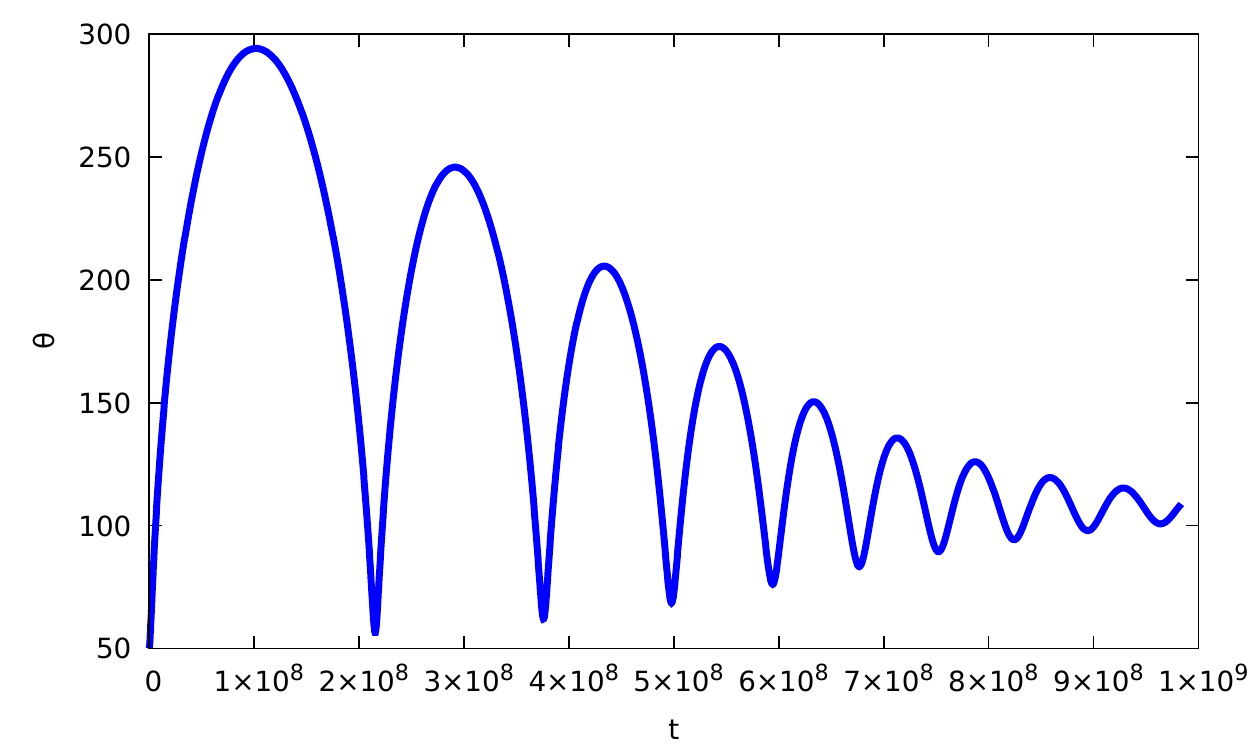}
	\includegraphics[width=0.48\textwidth]{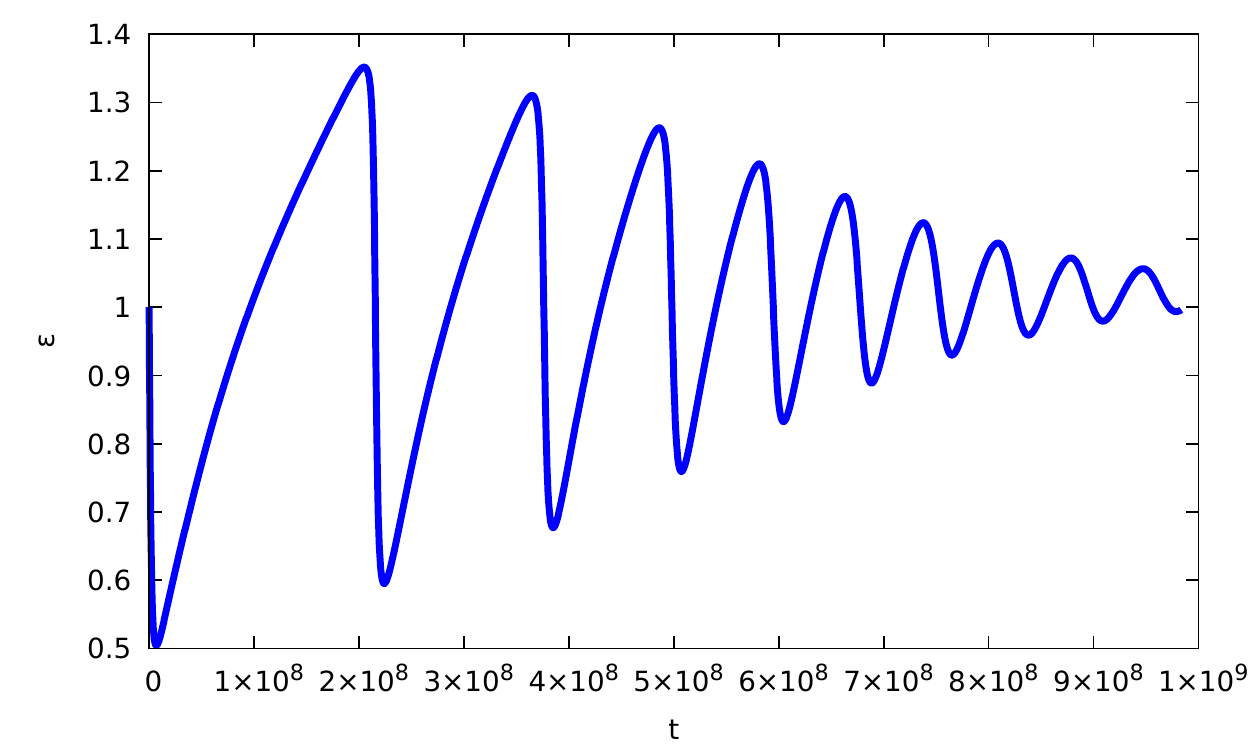}
	\includegraphics[width=0.48\textwidth]{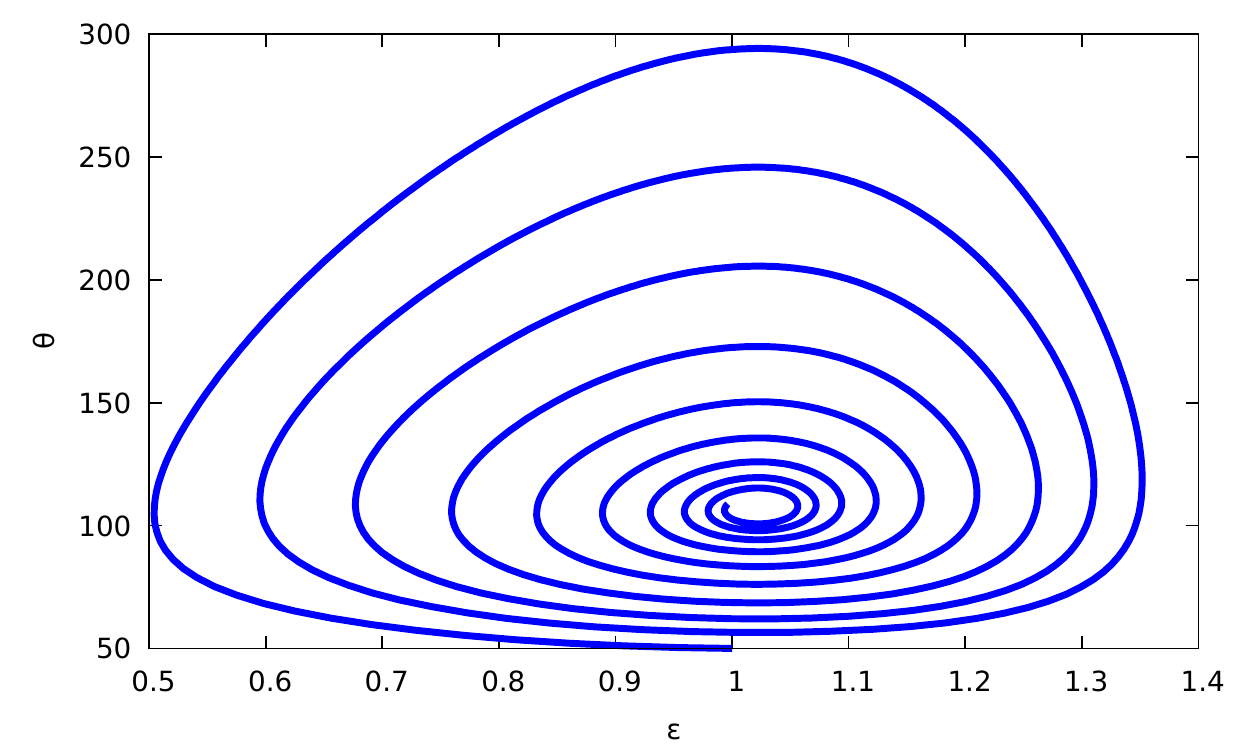}
	\caption{Evolution diagrams for a Betulia-shaped object with $A=0.5$. Time is in dimensionless units. The three panels show the thermal parameter $\theta$ versus time (top), the obliquity parameter $\varepsilon$ versus time (middle), and the phase diagram $\theta$ versus $\varepsilon$ (bottom). Settling to a stable equilibrium can be seen.}
	\label{fig:betulia}
\end{figure}

For this sake, we created a program, which simulates dynamical evolution of an asteroid. It solves Eqs. (\ref{dimensionless_evolution_equation1}) and (\ref{dimensionless_evolution_equation2}) with the right-hand-sides precisely computed for a given asteroid shape. The thermal model for $\tau_\varepsilon$ uses the Fourier algorithm described in Section \ref{sec:Obliquity component in terms of the YORP coefficients}.
Sample results of this simulation are shown in Figure \ref{fig:betulia}.

One can see a stable focal point. The asteroid starts its evolution far away from it, performs a number of oscillations with a decreasing amplitude, and eventually converges to the equilibrium. From Figure \ref{fig:betulia} it can be seen that the attraction basin around the focal point is large. Each asteroid whose shape permits such a stable equilibrium has a substantial probability of acquiring the initial conditions within this attraction basin, e. g. as a result of collision. Then the asteroid sinks to the equilibrium and resides there till the next collision or other major perturbation.

\section{Discussion}
\label{sec:Discussion}

The YORP effect is commonly evoked to explain some characteristic features of the asteroid population, in particular their distributions over the rotation rates and obliquities.

Concerning the obliquity distribution, the common perception is that the YORP effect tends to align the asteroid equatorial planes with their orbits, preferentially producing the obliquities $\varepsilon=0$ or $180^\circ$. Here we show that this conclusion is only partially true. It is indeed so when $C_\mathrm{cos}$ presents the dominant contribution to the obliquity component of YORP, as it can be seen in Figure \ref{fig:Csos-psysical-explanation}. Still from Figure \ref{fig:scales} we can see that $C_\mathrm{cos}$ wins over $C_\mathrm{sin}$ only in the minority of cases: about 1/4 cases for NEAs (when both $C_\mathrm{cos}/C_\mathrm{sin}$ and $\Gamma$ are above average -- each of these two independent conditions being satisfied in $\sim 50\%$ of cases) and even less for MBAs. In most other cases the equatorial and orbital planes become parallel or perpendicular with equal probabilities, as it is described in \cite{golubov19}.

As for the distribution of asteroid rotation rates under the influence of YORP, its simplified model has been proposed by \cite{pravec08}. The model assumed that each asteroid is created at zero rotation rate, then experiences a constant YORP acceleration all the way to the critical rotation rate, at which it gets disrupted. The authors successfully reproduced the observed flat distribution of asteroids over rotation rates with a sharp cutoff at the critical rotation rate around 10 turns per day.
The one-dimensionality of this model presents its most important flaw. The model only considers the change of the angular velocity $\omega$ but disregards the change of the obliquity $\varepsilon$. It is not a good approximation, as the axial component of YORP, $\tau_z$, is a function of the obliquity $\varepsilon$,
which in turn is not constant but also influenced by YORP.
A consistent study of evolution of asteroids must regard both variables, $\omega$ and $\varepsilon$, assume a realistic distribution of initial conditions for the asteroids starting their YORP cycles, and account for the possibility of the YORP equilibria.

Such equilibria can serve as attractors for asteroid evolution. After undergoing several disruptive YORP cycles and re-emerging from each of them with a new shape, an asteroid can eventually acquire such $C_\mathrm{sin}$ and $C_\mathrm{cos}$ to be locked in a stable rotation state. It would preserve this stable rotation until an external perturbation (collision, orbital change) either kicks it out or destroys its stability.
By its significance for the asteroid evolution, this equilibrium is similar to the ones previously discussed in literature \citep{golubov19,golubov16byorp,golubov18}, although caused by different physical mechanisms. It requires neither TYORP nor asteroid binarity, and thus in some sense it is the simplest of all types of equilibria.

Stability of these rotation states presents a major theoretical challenge. As in the simplest model one cannot determine whether the focal point is stable, a more sophisticated theory needs to be developed, accounting for the higher-order Fourier terms in Eqs. (\ref{T_eps_fit}) and (\ref{T_z_fit}). We will target the analytic and numeric study of these terms in the next article, thus making the simulation of the YORP evolution even more realistic.

The proposed model provides a compromise between accuracy and simplicity. It takes into account the thermal contribution to the YORP effect, but boils down the information about the asteroid shape to a few free parameters, allowing to keep track of their physical meaning and their individual impact on the simulation results. This model is similar to the one by \cite{golubov19}, and only slightly more complicated, but the inclusion of $C_\mathrm{cos}$ and the associated thermal model allows to bring in much new physics. The latter includes the new kind of stable equilibria and the preferential alignment of the equatorial and orbital planes at thermal parameters of the order of unity.
The new insight into the structure of asteroid families, dynamics of asteroid pairs, MBA-NEA asteroid transport can be obtained by conducting a rigorous simulation of the asteroid evolution similar to \cite{marzari20}. It should account for YORP, Yarkovsky and collisions, and such a simple but realistic model of YORP presents the cornerstone for such modeling of asteroid evolution.


\section*{Acknowledgements}

This work was partially funded by the National Research Foundation of Ukraine, project N2020.02/0371 ``Metallic asteroids: search for parent bodies of iron meteorites, sources of extraterrestrial resources''.

\appendix

\section{Methods for computation of the YORP effect}
\label{sec:Methods for computation of the YORP effect}

\begin{figure}
	\centering
	\includegraphics[width=0.48\textwidth]{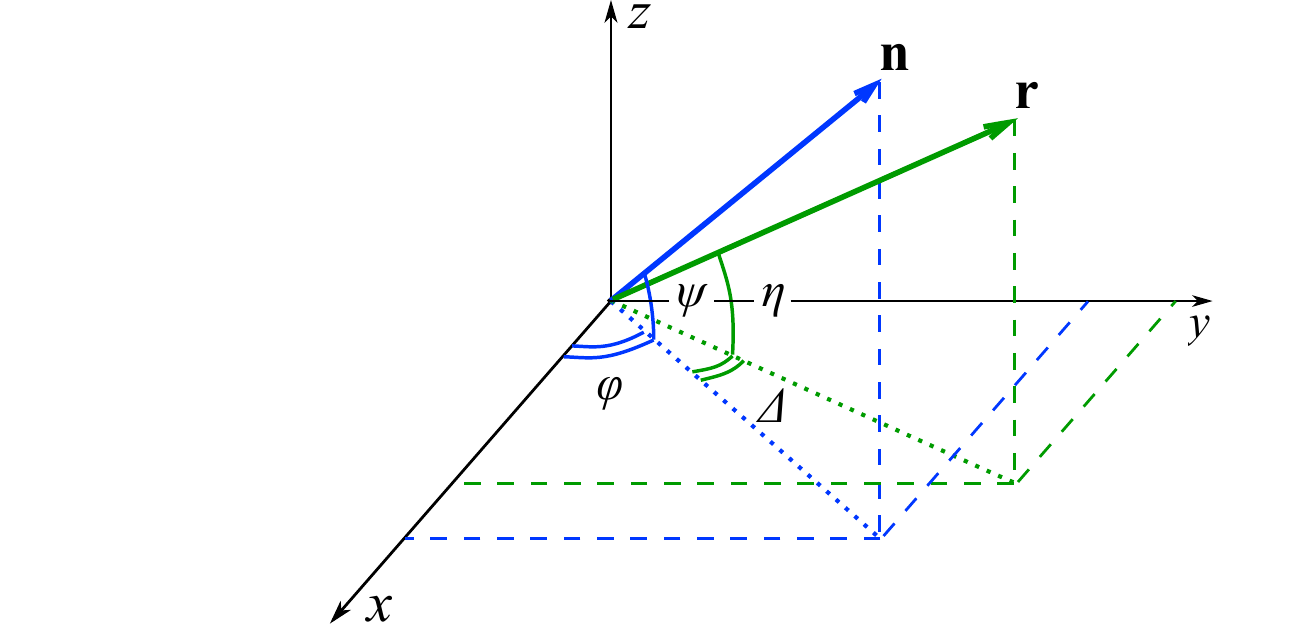}
	\caption{Orientation of the normal vector $\mathbf{n}$ and the radius vector $\mathbf{r}$ of the surface element with respect to the coordinate system.
		$\psi$ is the latitude of the surface element determined from its slope, $\eta$ is its latitude determined from the radius vector orientation.
		The angle $\phi$ between $Ox$ axis and the projection of $\mathbf{n}$ onto the equatorial plane $Oxy$ changes as the asteroid rotates, while the angle $\Delta$ between the projections of $\mathbf{r}$ and $\mathbf{n}$ remains constant.}
	\label{fig:geometry}
\end{figure}

Here we briefly review the YORP theory from \cite{golubov16nyorp}. The components of the non-dimensional YORP torque are expressed as integrals over the asteroid surface $S$ by the following equations:
\begin{align}
\tau_z = \frac{1}{R^3} \oint\limits_S \mathrm{d}S \, r\, \sin{\Delta}\cos{\eta}\cos{\psi} \, p^\alpha_z\left(\psi, \varepsilon\right),
\label{T_z_int}
\end{align}
\begin{align}
\label{T_epsilon_int}
\tau_{\varepsilon} = &-\frac{1}{R^3} \oint\limits_S \mathrm{d}S \, r\,\Big{(}\sin{\Delta}\cos{\eta}\sin{\psi} \times \\
\nonumber &\times \left(Ap^\alpha_\mathrm{sin}\left(\psi, \varepsilon\right)+(1-A)p^\tau_\mathrm{sin}\left(\psi, \varepsilon,\theta\right)\right) + \\
\nonumber & + \left(\cos{\psi} \sin{\eta} - \sin{\psi}\cos{\eta}\cos{\Delta}\right) \times \\
\nonumber & \times (1-A)p^\tau_\mathrm{cos}\left(\psi, \varepsilon,\theta\right)\Big{)}.
\end{align}
The angles $\psi$, $\eta$, $\Delta$ are defined by the orientation of a surface element on the asteroid, and explained in Figure \ref{fig:geometry}. $p^\alpha_z$, $p^\alpha_\mathrm{sin}$, $p^\tau_\mathrm{sin}$ and $p^\tau_\mathrm{cos}$ are the dimensionless YORP pressures. The former two of them are defined as follows,
\begin{align}
\label{p_a_z_simpl}
p^\alpha_z\left(\psi, \varepsilon\right) = &\frac{2}{3\pi^2} \int\limits^{\pi/2}_{-\pi/2} \mathrm{d}\phi \times \\
& \times \sqrt{1 - \left(\sin{\phi}\cos{\psi}\sin{\varepsilon} - \sin{\psi}\cos{\varepsilon}\right)^2}\ ,
\end{align}
\begin{align}
\label{p_a_sin_simpl}
p^\alpha_\mathrm{sin}\left(\psi, \varepsilon\right) = &\frac{2}{3\pi^2} \int\limits^{\pi/2}_{-\pi/2} \mathrm{d}\phi \sin{\phi} \times \\
\nonumber & \times \sqrt{1 - \left(\sin{\phi}\cos{\psi}\sin{\varepsilon} - \sin{\psi}\cos{\varepsilon}\right)^2}\ .
\end{align}
The latter two pressures, $p^\tau_\mathrm{sin}$ and $p^\tau_\mathrm{cos}$, are defined via the weighted averages of the dimensionless temperature, which in turn is defined by the heat conduction equation.
\begin{equation}
\label{p_t_sin_def}
p^\tau_\mathrm{sin}\left(\psi, \varepsilon,\theta\right) = \frac{1}{6\pi^2} \int\limits^{2\pi}_{0} \mathrm{d}\upsilon \int\limits^{2\pi}_{0} \mathrm{d}\phi\ \tau^4 \bigg|_{\zeta=0} \sin{\phi},
\end{equation}
\begin{equation}
\label{p_t_cos_def}
p^\tau_\mathrm{cos}\left(\psi, \varepsilon,\theta\right) = \frac{1}{6\pi^2} \int\limits^{2\pi}_{0} \mathrm{d}\upsilon \int\limits^{2\pi}_{0} \mathrm{d}\phi\ \tau^4 \bigg|_{\zeta=0} \cos{\phi}.
\end{equation}

With respect to both the arguments $\psi$ and $\varepsilon$, the function $p^\alpha_z$ is even, while the functions $p^\alpha_\mathrm{sin}$, $p^\tau_\mathrm{sin}$ and $p^\tau_\mathrm{cos}$ are odd:
\begin{align}
p^\alpha_z\left(\psi, \varepsilon\right)&=p^\alpha_z\left(-\psi, \varepsilon\right)=p^\alpha_z\left(\psi, -\varepsilon\right),\\
p^\alpha_\mathrm{sin}\left(\psi, \varepsilon\right)&=-p^\alpha_\mathrm{sin}\left(-\psi, \varepsilon\right)=-p^\alpha_\mathrm{sin}\left(\psi, -\varepsilon\right),\\
p^\tau_\mathrm{sin}\left(\psi, \varepsilon,\theta\right)&=-p^\tau_\mathrm{sin}\left(-\psi, \varepsilon,\theta\right)=-p^\tau_\mathrm{sin}\left(\psi, -\varepsilon,\theta\right),\\
p^\tau_\mathrm{cos}\left(\psi, \varepsilon,\theta\right)&=-p^\tau_\mathrm{cos}\left(-\psi, \varepsilon,\theta\right)=-p^\tau_\mathrm{cos}\left(\psi, -\varepsilon,\theta\right).
\end{align}

The functions $p^\tau_\mathrm{sin}$ and $p^\tau_\mathrm{cos}$ have the following limiting behavior,
\begin{align}
p^\tau_\mathrm{sin}\left(\psi, \varepsilon,0\right)=p^\alpha_\mathrm{sin}\left(\psi, \varepsilon\right),\quad
p^\tau_\mathrm{sin}\left(\psi, \varepsilon,\infty\right)=0, \\
p^\tau_\mathrm{cos}\left(\psi, \varepsilon,0\right)=0,\,\,\,\,\quad\quad\quad\quad
p^\tau_\mathrm{cos}\left(\psi, \varepsilon,\infty\right)=0.
\end{align}

Symmetries of the problem require that all functions  $p^\tau_\mathrm{sin}$ and $p^\tau_\mathrm{cos}$ vanish when either $\psi$ or $\varepsilon$ equals either 0 or $\pm\frac{\pi}{2}$. Moreover, the functions are odd with respect to both arguments $\psi$ and $\varepsilon$. This allows us to guess that qualitatively good approximations to $p^\tau_\mathrm{sin}(\psi, \varepsilon,\theta)$ and $p^\tau_\mathrm{cos}(\psi, \varepsilon,\theta)$ can be given by the expressions
\begin{equation}
\label{p_t_sin_approx}
p^\tau_\mathrm{sin}\left(\psi, \varepsilon,\theta\right) = p_\mathrm{sin}(\theta)\sin 2\varepsilon\sin 2\psi,
\end{equation}
\begin{equation}
\label{p_t_cos_approx}
p^\tau_\mathrm{cos}\left(\psi, \varepsilon,\theta\right) = p_\mathrm{cos}(\theta)\sin 2\varepsilon\sin 2\psi.
\end{equation}
This conjecture is also confirmed by Figures 3 and 4 in \cite{golubov16nyorp}. Therefore, we choose this form to interpolate $p^\tau_\mathrm{sin}$ and $p^\tau_\mathrm{cos}$.

\section{Analytic estimates for the YORP coefficient}
\label{sec:Analytic estimates for the YORP coefficient}
We define the two shape coefficients of the asteroid in the following way:
\begin{equation}
C_\mathrm{sin}=-\frac{1}{R^3}\oint\limits_S \mathrm{d}S \, r\sin{2\psi}\sin{\psi}\cos{\eta}\sin{\Delta},
\label{C_sin}
\end{equation}
\begin{eqnarray}
C_\mathrm{cos}=-\frac{1}{R^3}\oint\limits_S \mathrm{d}S \, r\sin{2\psi}\times\nonumber\\
\times\left(\cos{\psi} \sin{\eta} - \sin{\psi}\cos{\eta}\cos{\Delta}\right).
\label{C_cos}
\end{eqnarray}

To estimate $C_\mathrm{cos}$, we rewrite Eqn. (\ref{C_cos}): 
\begin{eqnarray}
C_\mathrm{cos}=-\frac{1}{R^3}\oint\limits_S \mathrm{d}S \, r\sin{2\psi}\times\nonumber\\
\times\left(\cos{\psi} \sin{\eta} - \sin{\psi}\cos{\eta}\right)\nonumber\\
-\frac{1}{R^3}\oint\limits_S \mathrm{d}S \, r\sin{2\psi}\left(1 - \cos{\Delta}\right)\sin{\psi}\cos{\eta}.
\label{A_cos_cos1}
\end{eqnarray}
To get an analytic estimate of $C_\mathrm{cos}$, we split tis equation into two parts, denoted as $C'_\mathrm{cos}$ and $C''_\mathrm{cos}$, and treat them separately: 
\begin{eqnarray}
C'_\mathrm{cos}=-\frac{1}{R^3}\oint\limits_S \mathrm{d}S \, r\sin{2\psi}\times\nonumber\\
\times\left(\cos{\psi} \sin{\eta} - \sin{\psi}\cos{\eta}\right),
\label{A_cos1}
\end{eqnarray}
\begin{eqnarray}
C''_\mathrm{cos}=-\frac{1}{R^3}\oint\limits_S \mathrm{d}S \, r\sin{2\psi}\left(1 - \cos{\Delta}\right)\sin{\psi}\cos{\eta}.
\label{A_cos11}
\end{eqnarray}

It is easy to see that for asteroids, whose shape is roughly spheroidal, $C'_\mathrm{cos}$ should be negative.
Indeed, for an oblate spheroid $\eta$ and $\psi$ have the same sign, and $\psi$ has a bigger absolute value. 
Then $\cos{\psi} \sin{\eta} - \sin{\psi}\cos{\eta}=\sin{(\eta-\psi)}$ has the sign opposite to $\sin{2\psi}$,
and the entire expression in Eqn. (\ref{A_cos1}) is negative.

On the other hand, $C''_\mathrm{cos}$ should always be positive, as the terms $1 - \cos{\Delta}$ and $\cos{\eta}$ are positive, while $\sin{2\psi}$ and $\sin{\psi}$ have the same sign.

\begin{figure}
	\centering
	\includegraphics[width=0.48\textwidth]{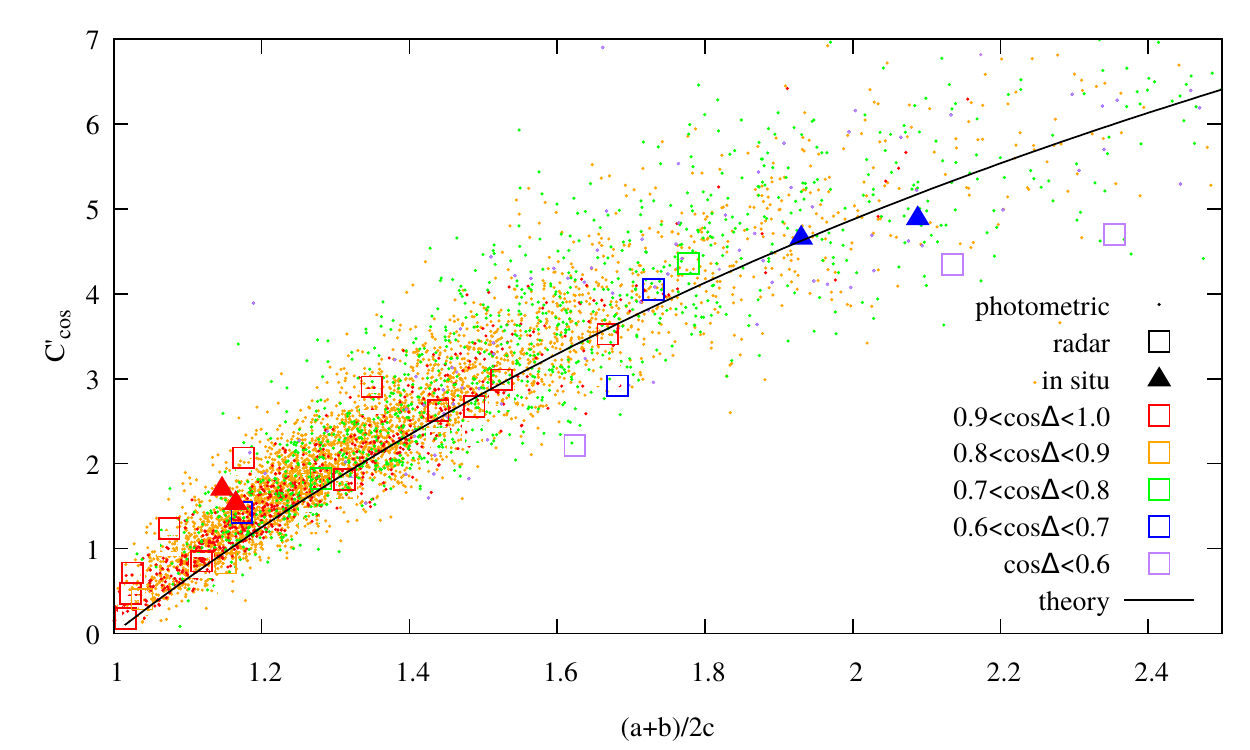}
	\caption{Dependence of $C'_\mathrm{cos}$ on ellipticity. The solid black curve is the prediction by Eqn. (\ref{A_cos1_ellipsoid}).}
	\label{fig:cos1_ellipticity}
\end{figure}

$C'_\mathrm{cos}$ can be easily evaluated for an oblate spheroid, which has $a=b>c$. It has the angle $\Delta=0$ (see Figure \ref{fig:geometry}), which much simplifies calculations. 
We parameterize the spheroid by representing it as a sphere anisortropically stretched in different directions, with $\alpha$ denoting the latitude on the initial sphere. Then the cylindric coordinates of a point on the spheroid can be expressed as $\rho=a\cos\alpha$, $z=c\sin\alpha$. The angles needed for the computation are $\tan\eta=z/\rho$, $\tan\psi=-d\rho/dz$, whereas the surface element is $dS=2\pi\rho\sqrt{d\rho^2+dz^2}$. 
Substituting these axpressions into Eqn. (\ref{A_cos1}) and computing the integral, we get
\begin{equation}
C'_\mathrm{cos}= 8\pi\left(\frac{2}{3}+\frac{1}{\frac{a^2}{c^2}-1}-\frac{\frac{a^2}{c^2}}{\left(\frac{a^2}{c^2}-1\right)^{3/2}}\arctan\sqrt{\frac{a^2}{c^2}-1}\right).
\label{A_cos1_ellipsoid}
\end{equation}
The Taylor decomposition of this expression gives the following approximation valid for $|a-c|\ll c$:
\begin{equation}
C'_\mathrm{cos}\approx\frac{32\pi}{15}\left(\frac{a}{c}-1\right).
\label{A_cos1_ellipsoid_linear}
\end{equation}
For a triaxial ellipsoid the integral cannot be computed in a closed form, but as a simple estimate we can substitute $a$ in the former expressions with the mean equatorial radius $(a+b)/2$.
Equation (\ref{A_cos1_ellipsoid}) in this approximation is plotted in Figure \ref{fig:cos1_ellipticity} with a solid black line. It agrees well with the results of simulations for different asteroid shapes. No dependence is seen between $C'_\mathrm{cos}$ and the asteroid's surface roughness, which is color coded. 

\begin{figure}
	\centering
	\includegraphics[width=0.48\textwidth]{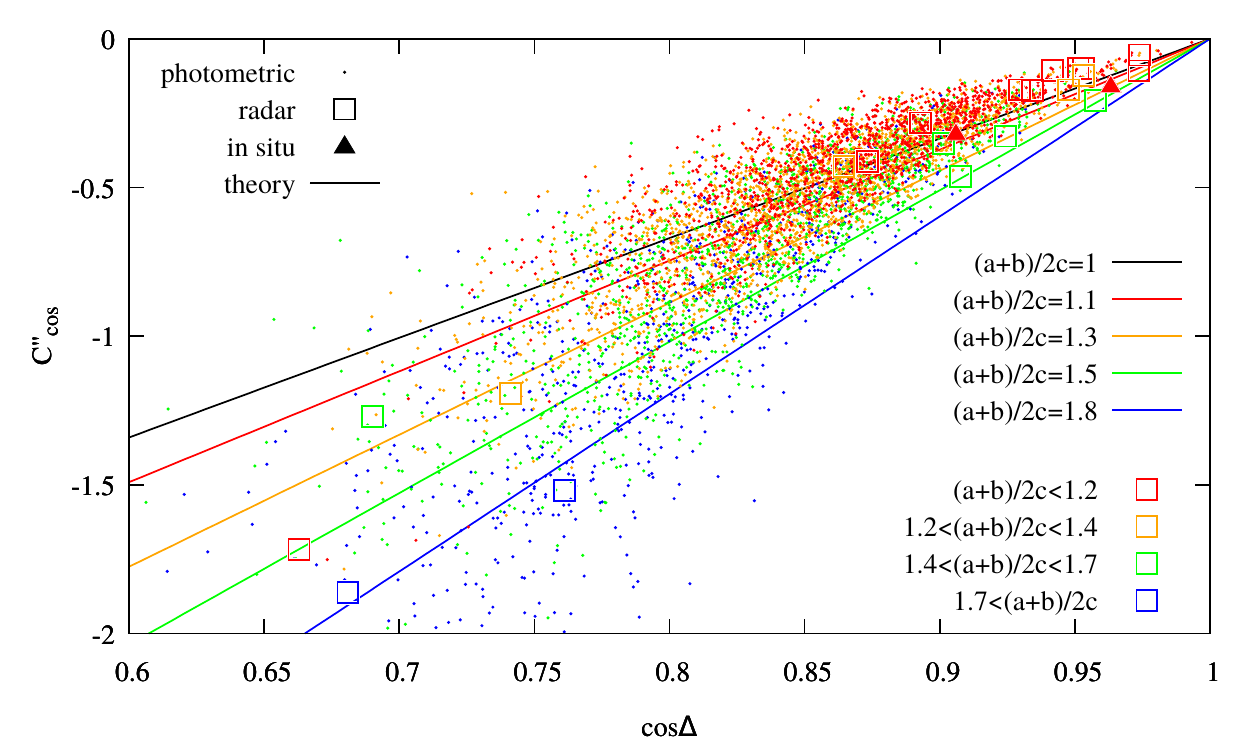}
	\caption{Dependence of $C''_\mathrm{cos}$ on the surface roughness. The solid black curve is the prediction by Eqn. (\ref{A_cos11_ellipsoid}).}
	\label{fig:cos2_Delta}
\end{figure}

\begin{figure}
	\centering
	\includegraphics[width=0.48\textwidth]{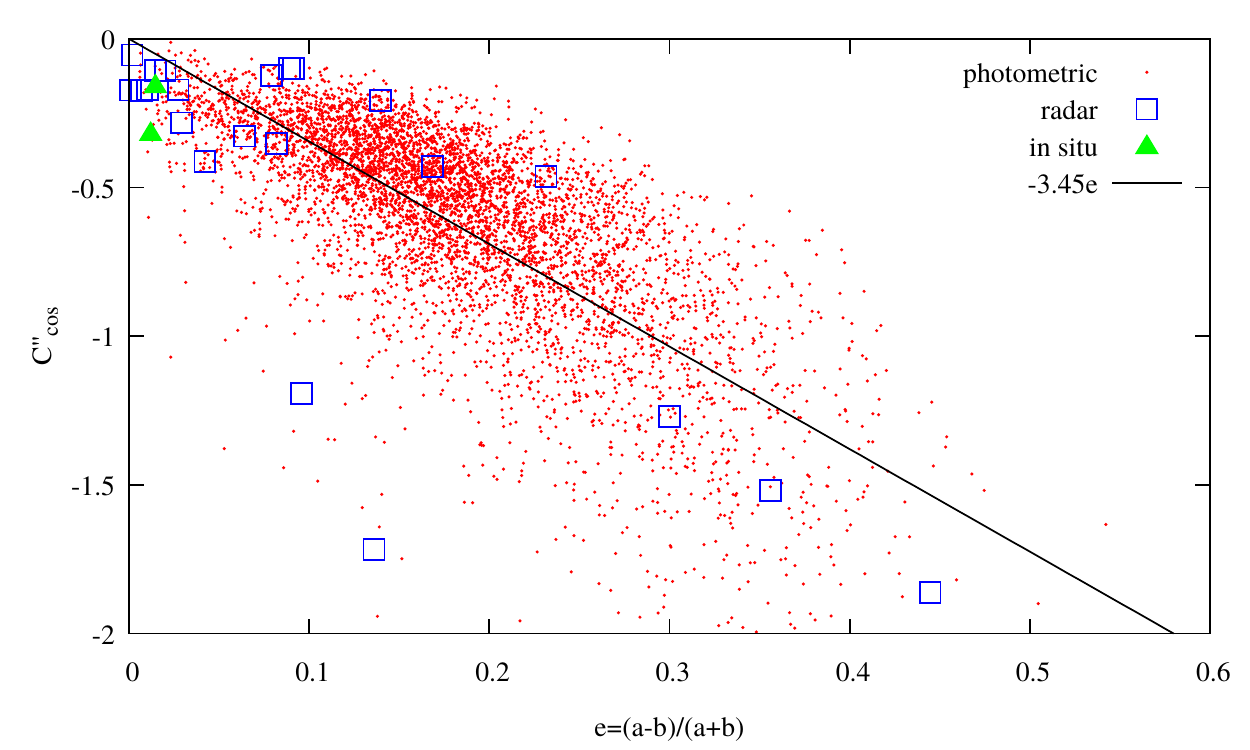}
	\caption{Dependence of $C''_\mathrm{cos}$ on the ellipticity of  asteroid equator. The solid black curve is the linear fit to the photometric data.}
	\label{fig:cos2_ellipticity}
\end{figure}

A simple estimate for $C''_\mathrm{cos}$ can be obtained in the following manner. Let $\bar{\Delta}$ to be some effective value of $\Delta$. We assume it roughly constant, and take it out of the integral. The remaining integral is evaluated in the same way as Eqn. (\ref{A_cos1_ellipsoid}), leading to a similar expression:
\begin{equation}
C''_\mathrm{cos}=-\frac{8\pi\left(1-\cos{\bar{\Delta}}\right)}{1-\frac{c^2}{a^2}}\left(\frac{2}{3}+\frac{1}{\frac{a^2}{c^2}-1}-\frac{\frac{a^2}{c^2}}{\left(\frac{a^2}{c^2}-1\right)^{3/2}}\arctan\sqrt{\frac{a^2}{c^2}-1}\right).
\label{A_cos11_ellipsoid}
\end{equation}
Again, we use this equation as an approximation for triaxial ellipsoids, by substituting $(a+b)/2$ instead of $a$. This ignores the fact that the difference between $a$ and $b$ causes a complicated correlation between $\Delta$ and the position on the asteroid. The Taylor approximation to Eqn. (\ref{A_cos11_ellipsoid}) in the case of $|a-c|\ll c$ gives
\begin{equation}
C''_\mathrm{cos}\approx-\frac{16\pi}{15}\left(1-\cos{\bar{\Delta}}\right).
\label{A_cos11_ellipsoid_linear}
\end{equation}
This equation is illustrated in Figure \ref{fig:cos2_Delta}. We see a good agreement between Eqn.(\ref{A_cos11_ellipsoid}) and the simulations for the shape models.

In Figure \ref{fig:cos2_ellipticity}, $C''_\mathrm{cos}$ is studied as a function of the ellipticity of the asteroid's equator, $e=(a-b)/(a+b)$. The figure shows a negative trend, which can be fitted by a linear dependence. Still, the trend is worse than in Figure \ref{fig:cos2_Delta}. This is natural, as high ellipticity of the equator implies a large average deviation $\Delta$ between the equatorial projections of the radius-vector and the normal vector. 

\begin{figure}
	\centering
	\includegraphics[width=0.48\textwidth]{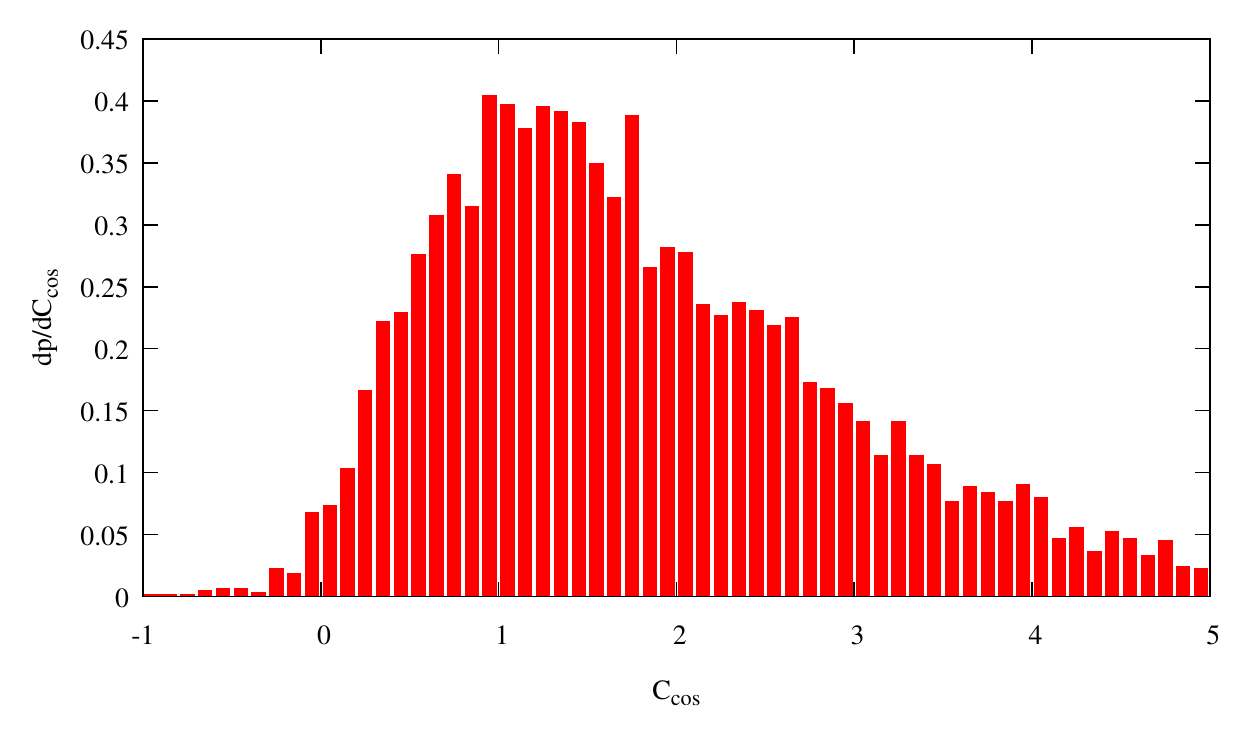}
	\caption{Histogram, showing the distribution of asteroids from the DAMIT database over $C_\mathrm{cos}$.}
	\label{fig:histogram_cos}
\end{figure}

The distribution of asteroids from the DAMIT database over the coefficient $C_\mathrm{cos}$ is shown in Figure \ref{fig:histogram_cos}.
We see once again that in almost all the cases this coefficient is positive.

\end{document}